\documentclass[a4paper,11pt]{article}
\usepackage{jheppub} 
\usepackage{lineno}
\usepackage{bbold}
\usepackage{xcolor,multirow,makecell}
\usepackage[normalem]{ulem}
\usepackage{subfig}
\usepackage{xspace}
\usepackage[countmax]{subfloat}

\usepackage{booktabs}
\usepackage[normalem]{ulem}

\title{\boldmath 
Prospects for sub-EW-scale ALP searches via $\gamma+b\bar{b}$ signatures at the LHC using jet substructure techniques 
}

\author[a]{Amit Adhikary,}
\author[a]{Aoife Bharucha,}
\author[b]{Lorenzo Feligioni}
\author[c]{and Michele Frigerio}

\affiliation[a]{Aix Marseille Univ, Universit\'{e} de Toulon, CNRS, CPT, IPhU, Marseille, France}
\affiliation[b]{Aix Marseille Univ, CNRS/IN2P3, 
CPPM, Marseille, France}
\affiliation[c]{Laboratoire Charles Coulomb (L2C), University of Montpellier and CNRS, Montpellier, France}

\emailAdd{amit.adhikary@cpt.univ-mrs.fr}

\abstract{
The current Large Hadron Collider (LHC) data show no clear indication of new physics and only incremental improvements are anticipated at the energy frontier in the near future. However, while the focus of the LHC has been on constraining TeV scale physics, new particles could still be hiding below the electroweak scale. In order to obtain sensitivity to a new light boson with couplings to SM fermions, a potentially promising decay channel, for resonances with mass $\gtrsim {\cal O}(10)$ GeV, would be the decay to $b\bar b$ pairs. The measurement of such signatures is challenging due to the trigger requirements at the LHC. In this work, we explore the LHC sensitivity to a light pseudoscalar, or axion-like particle (ALP), in the $b\bar{b}$ final state with an associated photon, using jet substructure techniques, in the mass range between 10 GeV and 100 GeV. We obtain projected exclusions on the ALP-fermion coupling in a region of phase space which has not so far been probed by direct searches. We further discuss the impact that lower trigger thresholds may have on the LHC reach. }

\begin{document}
\maketitle
\flushbottom

\section{Introduction}
\label{sec:intro}

The excellent performance of the LHC in terms of delivered luminosity has allowed the ATLAS and CMS experiments to set stringent limits on new particle masses well beyond the electroweak (EW) scale, thus worsening the naturalness problem.
However, it is still possible that the new physics is hidden at lower energies, being weakly coupled to the known Standard Model (SM) particles, such that any sign of its production are swamped by the SM background.
Current resonance searches at the LHC leave an area of opportunity roughly below the EW scale, which could potentially be accessed via the $b\bar{b}$ final state.
Light scalars and pseudoscalars are particularly promising candidates, with possibly large branching ratios to $b\bar{b}$ if their masses are above 10 GeV, as further discussed in Section~\ref{sec:model}.
The resonances may be identified with well-motivated axion-like particles (ALPs)~\cite{Jaeckel:2010ni,DiLuzio:2020wdo,Brivio:2017ije,Bauer:2017ris,ParticleDataGroup:2024cfk}. They could be mediators between the SM and dark matter (DM)~\cite{XiaoyongChu_2012,Boehm:2014hva,Arina:2014yna,Hambye:2019dwd,Bharucha:2022lty}. 
Such spin-zero particles with masses below the EW scale could further arise in supersymmetric models, such as the next-to-minimal supersymmetric SM \cite{Ellwanger:2009dp}
or models with an approximate $R$-symmetry \cite{Bellazzini:2017neg}, or
composite Higgs models with additional Nambu-Goldstone bosons \cite{Gripaios:2009pe,Frigerio:2012uc,Cacciapaglia:2019bqz,BuarqueFranzosi:2021kky,Elander:2020nyd}.

Previous searches targeting low mass hadronic resonances have been carried out by the ATLAS and CMS collaborations. From the experimental point of view, the challenge starts already at the trigger level, where stringent requirements on hadronic jets transverse momentum ($p_{\rm T}$) are usually needed in order to cope with the bandwidth limitations of the LHC experiments data acquisition system. ATLAS searched for singly produced $q\bar{q}$ resolved resonances in 29.3 fb$^{-1}$ of proton-proton collision at a center-of-mass energy $\sqrt{s}=13$~TeV, using Trigger Level Analysis~\cite{ATLAS:2018qto} (TLA), a technique that allows to lower the trigger threshold by recording only the information calculated by the jet trigger algorithms in the High Level Trigger. Making use of a single jet trigger with a $p_{\rm T}$ threshold of 185~GeV, the TLA sets a limit on an axial-vector DM mediator with couplings to SM quarks and mass above 450~GeV. 

The production of light resonances is often investigated via associate production with other particles, such as high energetic photons or quark and gluons from initial state radiation (ISR), less overwhelmed by the SM processes and consequently less challenging to trigger on, though presenting significantly smaller and possibly more model-dependent production rates.  
CMS searched for boosted light resonances decaying into collimated quark pairs, recoiling from high-$p_{\rm T}$  quark/gluon radiation, using a data sample of  41.1 fb$^{-1}$ of proton-proton collision at a center-of-mass energy of $\sqrt{s}=13$~TeV~\cite{CMS:2019emo}. 
These resonances are looked for in single large radius jet, reconstructed with the anti-$k_{\textrm{T}}$ clustering algorithm~\cite{Cacciari:2008gp} of radius R=0.8 (AK8) and distinguished from the dominant QCD background by exploiting the  jet substructure. 
For this search CMS uses a combination of triggers based on the  scalar sum of the jet $p_{\rm T}$s ($H_{\rm T}$) and on leading jet $p_{\rm T}$ thresholds. The offline  leading AK8 jet is required to have $p_{\rm T}>$ 525 GeV, this condition ensures that the trigger is fully efficient with respect to the offline selection.  
By identifying the leading jet as produced by the resonance decay products, CMS is able to set a limit on lepto-phobic $Z'$ with masses from 50 to 250 GeV. 
CMS also searched for low mass resonances as an excess arising in the invariant mass spectrum of two-pronged substructure within an AK8 jet recoiling from a high energetic photon~\cite{CMS:2019xai}. These events are selected in 41.1 fb$^{-1}$ of proton-proton collisions at a center-of-mass energy of $\sqrt{s}=13$~TeV using a single photon trigger with $p_{\rm T}$ threshold of 175 GeV which translates to a 200 GeV $p_{\rm T}$ requirement for the offline reconstructed photon to be at plateau of the trigger efficiency.  
Thanks to the cleaner photon signature, the leading AK8 jet $p_{\rm T}$ requirement, equivalent to the photon $p_{\rm T}$, is lower than the one used in hadronic ISR-based analyses, as a consequence upper limits are placed on the quark coupling for masses as low as 10 GeV.

ATLAS as well performed a search for low mass resonances produced in association with either a photon or a high energetic ISR jet, using an integrated luminosity of  36.1 fb$^{-1}$ at a center-of-mass energy of $\sqrt{s}=13$ TeV~\cite{ATLAS:2018hbc}. 
The trigger requires either an isolated photon with transverse energy $E_{\rm T} >$ 140 GeV or an anti-$k_{\textrm{T}}$ jet with radius parameter R = 0.4 (AK4) and $p_{\rm T} >$ 280 GeV. The resonance is searched as an excess in the distribution of the large-radius jet masses, reconstructed with the anti-$k_{\textrm{T}}$ algorithm of radius R=1.0 (AK1.0) and requiring $p_{\rm T} > $ 200 GeV, in the mass range between 100 GeV and 220 GeV.  
ATLAS searched for resonance decaying into $b\bar{b}$ pairs using 79.8 fb$^{-1}$ of data recorded at a center-of-mass energy of $\sqrt{s}=13$ TeV with a single photon trigger and 76.6 fb$^{-1}$ with a photon plus jet trigger~\cite{ATLAS:2019itm}. The resonance is considered to recoil from the photon and to decay into two distinct reconstructed AK4 jets with transverse momentum $p_{\rm T}>$ 25 GeV jets which  contain $b$-hadrons ($b$-jets) are identified using specialised algorithms ($b$-tagged). Given the sizable angular separation between individual jets the mass range probed for the presence of a $Z'$ axial-vector dark-matter mediator range from 169 to 1100 GeV.
Note that the expected sensitivity for the $b$-tagged analysis is better than the flavour-inclusive, non $b$-tagged analysis, even for $Z'$ models where the $Z'$ equally decay into all quark flavours. 

CMS also searched for a low mass resonance decaying into $b\bar{b}$ pairs recoiling from an ISR jet in events recorded at a center-of-mass energy of $\sqrt{s}=13$ TeV corresponding to a luminosity of 35.9 fb$^{-1}$ of data~\cite{CMS:2018pwl} using a combination of several triggers whose full efficiency condition requires the $p_{\rm T}$ of the large-radius jet used for reconstructing the resonance mass to be larger than 450 GeV. 
A dedicated double-$b$-tagger is used to select jets likely to originate from two $b$-quarks. No significant excess above the SM prediction is  observed for signal masses between 50–350 GeV and limits were set in this region.

In 2030 the High Luminosity (HL-LHC) phase of the LHC will begin~\cite{ATL-PHYS-PUB-2019-005,Collaboration:2650976}, which will allow the ATLAS and CMS experiments to collect an integrated luminosity of 3000 fb$^{-1}$ of proton-proton collisions at a center-of-mass energy of $\sqrt{s}=14$~TeV. Right before HL-HLC starts, both detectors will undergo several upgrades that will result in improved reconstruction and pile-up suppression capabilities. This work aims at optimizing the analysis strategy in order to maximize the experimental sensitivity to detect pseudoscalar resonances decaying to $b\bar{b}$ pairs at the end of HL-HLC.
We perform our analysis focusing on the sensitivity for resonances with mass values ranging from 10 GeV to 100 GeV. In this mass range, we adopt a strategy based on triggering on a high-$p_{\rm T}$ photon and we look for a recoiling large-radius jet that contains collimated $b\bar{b}$ pairs. As discussed before, this is an optimal strategy, given the lower achievable photon threshold at trigger level with respect to choosing ISR jets, it can probe efficiently masses $\gtrsim 10$ GeV. Moreover, the $b\bar{b}$ final state is the dominant decay channel for spin-0 particles with Yukawa-like couplings to fermions and masses below the top-quark mass. To be noticed that the analysis proposed is an extension of the current landscape of light $b\bar{b}$ resonance searches. 

The structure of the paper is organised as follows. Section~\ref{sec:model} presents the model of interest, which features a light pseudoscalar with Yukawa-like coupling structure to fermions. In Section~\ref{sec:collider}, we present a collider analysis for the pseudoscalar decaying to $b\bar{b}$ final state. Section~\ref{sec:event_sim} details the simulation of signal and background events. The event selection criteria is presented in Section~\ref{sec:event_sel} together with the description of jet substructure observables used to reject background events. In Section~\ref{sec:di-btag}, the emulation of $b$-tagging algorithms are presented on simulated events, together with the  labelling scheme used to categorise the AK1.0 jets according to the \textit{flavour}. A fit is performed on the signal mass distribution in Section~\ref{sec:Nevents}, and the signal efficiencies and background yields are estimated after the analysis requirements. Section~\ref{sec:results} presents the sensitivity of the analysis to exclude a $b\bar{b}$ resonance at the HL-LHC, including a modified trigger requirement to further enhance  the sensitivity in Section~\ref{sec:low_pho_pt}. 
The importance of the results are interpreted with respect to the existing bounds in Section~\ref{sec:summ_result}. 
Finally, we summarise our findings in Section~\ref{sec:conclusion}, draw conclusions and outline future directions.

\section{Light pseudoscalar: model and motivations}
\label{sec:model}

The model we study consists of a light pseudoscalar, also referred as axion-like particle (ALP).
We consider this particle $a$ to couple only to SM fermions, and assume these couplings to be proportional to the fermion mass,
\begin{align}
\mathcal{L}_{\rm ALP} \supset &\;\frac{1}{2} \partial_\mu a \partial^\mu a - \frac{1}{2}m^{2}_{a} a^2 + i\sum_f  g_{\rm aff} m_f \,a\bar f \gamma_5 f  \,, \label{eq:L}
\end{align}
where $m_a$ is the mass of the ALP, and $g_{\rm aff}$ is a real, dimensionful parameter, $g_{\rm aff}=\frac{C_f}{f_a}$, with $C_f$ dimensionless and $f_a$ an energy scale, typically the spontaneous-symmetry-breaking scale associated with the ALP.
Assuming $C_f\lesssim \mathcal{O}$(1)  and $f_a$ is greater than the EW symmetry breaking scale, one obtains an approximate theoretical upper limit of $g_{\rm aff}\lesssim 0.01$ GeV$^{-1}$.

We assume $C_f$ to be universal for all SM fermions. 
This assumption is motivated by the following argument
\cite{Brivio:2017ije,Quevillon:2019zrd,Bauer:2020jbp,Arias-Aragon:2022iwl}. 
If the ALP interactions respect a shift symmetry, $a\to a + c$ (as expected for Nambu-Goldstone bosons), and if one requires a principle of minimal flavour violation (to circumvent experimental constraints on flavour and CP violation), then
the ALP couplings to SM quarks have to be proportional to the quark Yukawa couplings,
\begin{equation}
{\mathcal L}_{\rm ALP}\supset i \frac{a}{f_a} 
\left(C_d \overline{q_L}Y_d d_R H + C_u  \overline{q_L}Y_u u_R \tilde{H} +h.c.\right)\,,
\end{equation}
with $C_{d,u}$ real constants, up to corrections involving higher powers of $Y_{d,u}$.
Now, replacing the Higgs doublet $H$ by its vacuum expectation value, and diagonalising the quark mass matrices, 
one is led to Eq.~\eqref{eq:L}, with $C_d=C_s=C_b$ and $C_u=C_c=C_t$.
In our analysis we will take for simplicity $C_d=C_u$ but,
as we are interested in ALP decays to $b$ quarks, one could suppress $C_u$, to avoid potential, indirect constraints on
ALP couplings to the up quark, as obtained in Ref.~\cite{Biekotter:2023mpd} (see discussion in Section~\ref{sec:summ_result}).
The Yukawa-like couplings of the ALPs to fermions imply that for ALPs with masses between twice the bottom mass and twice the top mass, $2 m_b<m_a<2 m_t$, the dominant decay is $a\to b\bar{b}$.
This motivates LHC searches for $\mathcal{O}(10-100)$ GeV resonances decaying to $b\bar{b}$ final states.
In order to trigger on this final state, we require an initial state photon, as further discussed in Section~\ref{sec:collider}.

Note that such an ALP could also couple to DM, as studied for the case of fermionic DM in e.g.~Refs.~\cite{Bharucha:2022lty,Ghosh:2023tyz}. 
The Lagrangian takes the form
\begin{align}
\mathcal{L} =&\;{\cal L}_{\rm SM}+{\mathcal L}_{\rm ALP}+\bar\chi (i\partial\!\!\!\slash-m_\chi)\chi + i\frac{ C_\chi}{f_a}\,m_\chi\, a\bar \chi \gamma_5 \chi\,, \label{eq:lagrangian}
\end{align}
where the DM particle $\chi$ of mass $m_{\chi}$ couples to the ALP via the coupling $C_\chi$.
Depending on the couplings of the ALP to DM and to the SM, the production mechanism changes.
In freeze-out, one has large enough ALP couplings such that equilibrium is obtained between the hidden sector particles and SM, whereas in freeze-in, the couplings are small, such that DM-SM equilibrium is never achieved.
In both cases, the relic density may depend on the ALP to DM thermal cross-section or the SM to DM thermal cross-section, depending on which of these are dominant.
Considering the $m_a$ - $g_{\rm aff}$ plane, it was found in Ref.~\cite{Bharucha:2022lty} that a large range of parameter space, where the correct relic density via freeze-in can be obtained, remains allowed, particularly for values of $m_a$ above $\mathcal{O}(10)$ GeV.
Note that for the values of $g_{\rm aff}$ close to the theoretical upper limit of $g_{\rm aff}\lesssim 0.01$ GeV$^{-1}$ (which could be probed at colliders) dark matter production via fermion scattering is relevant.
It was seen in Ref.~\cite{Bharucha:2022lty} that in this region there is a large hierarchy between $g_{\rm aff}$ and $g_{a\chi\chi}$, which would need to be explained via further model building.

The sensitivity study performed in this work is presented in terms of the above-discussed ALP model.
However, pseudoscalars with masses below the electroweak scale and couplings to fermions are also motivated by theories addressing the stability of the electroweak scale, such as Higgs compositeness.
In fact, a composite Higgs boson requires the introduction of a new strongly-coupled sector close to the TeV scale, whose spectrum contains several composite resonances, including SM singlets that may be lighter than the Higgs itself. 
In this context, a particularly well-motivated target for low mass resonance searches  is a pseudo Nambu-Goldstone boson $a$, analogous to a QCD pion, whose small mass arises radiatively, via shift-symmetry breaking couplings. 
Indeed, the couplings of the $a$ to $b$-quarks induce $m_a\gtrsim 10$ GeV~\cite{Gripaios:2009pe,Frigerio:2012uc} at one loop, as desired. 
Note that here we are referring to additional ALP couplings, which break the Nambu-Goldstone shift symmetry, thus inducing a non-trivial potential.
Such a composite pseudoscalar has been searched for at colliders, but mostly using large-$p_{\rm T}$ signatures which require $m_a$ above the electroweak scale \cite{Cacciapaglia:2019bqz}.
Interesting studies to probe $m_a$ well below the electroweak scale at LHCb can be found in Refs.~\cite{CidVidal:2018blh,BuarqueFranzosi:2021kky}, which exploit decay channels other than 
$b\bar b$.

\section{Collider Analysis}
\label{sec:collider}

\begin{figure}[!tb]
\begin{center}
\subfloat[]{\label{fig:sig}
\includegraphics[width=.25\textwidth]{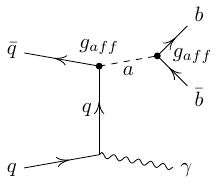}
}\qquad
\subfloat[]{\label{fig:bkg1}
\includegraphics[width=.25\textwidth]{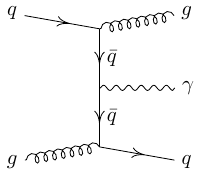}
}\qquad
\subfloat[]{\label{fig:bkg2}
\includegraphics[width=.25\textwidth]{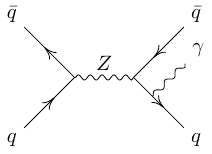}
}
\end{center}
\caption{(a) The Feynman diagram for the signal process, $pp\to a(\to b\bar{b})\gamma$, where the ALP-fermion couplings, $g_{\rm aff}$, are shown, and one representative Feynman diagram for the (b) non-resonant and (c) resonant  background, $pp\to jj\gamma$.}
\label{fig:FD}
\end{figure}

This work focuses on the search of a light pseudoscalar, in the mass range between 10 GeV and 100 GeV, decaying to $b\bar{b}$ pairs as predicted in the ALP model described in Section~\ref{sec:model}, at the LHC.
In order to reconstruct its decay products, the low mass resonance is required to be in a boosted regime, where the produced $b$-hadrons from the ALP are 
collimated.
Kinematically this can be achieved by demanding that the ALP is produced together with a high energetic recoiling object, either a photon or an ISR jet.
The signal in the current analysis encompasses a high $p_{\rm T}$ isolated photon produced back-to-back to a large radius jet having two prong jet substructure from hadronisation objects of two $b$-quarks. The Feynman diagram of the hard process for signal production, $pp\to a(\to b\bar{b})\gamma$, is shown in Fig.~\ref{fig:FD} (a)\footnote{We showed only the $q\bar q$ initial state, because the putative gluon fusion diagram for $\gamma +$ALP production vanishes. Indeed, the ALP couplings to quarks, $C_q$, allow for gluon-gluon fusion to produce an ALP via a triangle loop of quarks, an effect dominated by the top quark loop. However, we are interested in the ALP production associated with a photon, which rather involves a box loop of quarks, and the latter vanishes as a consequence of the generalised Furry's theorem \cite{Nishijima1}.}.

The dominant background arises from multijet production in association with a photon. In Fig.~\ref{fig:FD} (b), we show one of the Feynman diagrams for the $\gamma+$multijet process. Subdominant resonant backgrounds originate from $W+\gamma$ and $Z+\gamma$ processes (see Fig.~\ref{fig:FD} (c)). 
Since modeling uncertainties could have a large impact on the results, simulated background events are only used here to validate the strategy of the data-driven background estimation. 
Despite a plethora of next-to-leading order (NLO) generators~\cite{Catani:2002ny,Sherpa:2024mfk} being available and currently used by both ATLAS and CMS collaborations to compare with data in specific phase-space regions~\cite{ATLAS:2016fta,ATLAS:2023yrt}, the LHC experiments still rely on leading order (LO) generators to provide large samples of $\gamma$ + jets events to validate data-driven techniques~\cite{CMS:2019xai,CMS:2017dcz}.
In the following subsections, we discuss the generation of signal and background processes, the physics object used in the analysis, kinematic observables, the event selection criteria, and finally the details of the $b$-tagging simulation. The section ends with the estimation of the signal and background yields.

\subsection{Event simulation}
\label{sec:event_sim}

Signal and SM background processes are simulated using Monte Carlo (MC) event generators.
The $\gamma +$ALP signal process $pp\to a(\to b\bar{b})\gamma$ is generated using \texttt{MadGraph5\_aMC@NLO}~\cite{Alwall:2014hca} at LO. The $\gamma$+multijet background  process, which include both non-resonant and resonant $W/Z+\gamma$ processes, is simulated using LO generator \texttt{Pythia8}~\cite{Sjostrand:2014zea}. Both signal and background are generated in the 5F scheme where proton contains gluon, light- and $b$-quarks. 
The center-of-mass energy considered for the proton-proton collisions is  $\sqrt{s}=14$ TeV. To simulate the showering and hadronisation effects, the generated events are interfaced with \texttt{Pythia8},
with the \texttt{A14} tune~\cite{ATL-PHYS-PUB-2014-021}, and the parton distribution function (PDF) set \texttt{NNPDF2.3NLO}~\cite{NNPDF:2014otw} is used. Finally, the detector simulation is performed with \texttt{Delphes-3.5.0}~\cite{deFavereau:2013fsa} using the default HL-LHC ATLAS analysis card.  We apply specific requirements for the generation of the signal and background events to improve the statistics in the phase space relevant to the proposed analysis. The inclusive cross-sections and generation-level cut efficiencies for the $pp\to a(\to b\bar{b})\gamma$ signal process is shown in Table~\ref{tab:cs_cuts}. 

In order to produce on-shell light pseudoscalar, decaying into two $b$-hadrons, we restrict the lowest ALP mass to $m_a> 2m_{\rm B^0/B^\pm}\simeq 10.6$ GeV. Specifically, we choose the following benchmark masses for the ALP, $m_a=$ 12 GeV and then from 20 GeV to 100 GeV in steps of 10 GeV. The ALP-fermion coupling, $g_{\mathrm{aff}}$ is fixed to $0.1$ GeV$^{-1}$ and the total decay width of the ALP for each benchmark mass is taken from Ref.~\cite{Bharucha:2022lty}. 

\renewcommand{\arraystretch}{1.3}
\begin{table}[tb!]
\centering
\begin{tabular}{c| c |c}
\toprule
\multicolumn{3}{c}{$pp\to a(\to b\bar{b})\gamma$ }\\\hline
\makecell{$m_a$ \(\rm [GeV]\) } & \makecell{$\sigma$  ($g_{\mathrm{aff}}=0.1$ GeV$^{-1}$) \(\rm [pb]\)
} & \makecell{$\epsilon$ ($\times 10^{-5}$)} \\\hline
12 & 15264 $\pm$ 113 & 0.07 \\
20 & 9727 $\pm$ 82 & 0.13\\
30 & 4690 $\pm$ 43 & 0.27\\
40 & 2454 $\pm$ 22 & 0.48\\
50 & 1343 $\pm$ 13 & 0.79\\
60 & 779 $\pm$ 7 & 1.22\\
70 & 472 $\pm$ 5 & 1.78\\
80 & 294 $\pm$ 3 & 2.49\\
90 & 191 $\pm$ 2 & 3.34\\
100 & 128 $\pm$ 1 & 4.35\\\toprule
\end{tabular}
\caption{The inclusive cross-section ($\sigma$) for $pp\to a(\to b\bar{b})\gamma$ production, and generation cut efficiency ($\epsilon$) which is defined as the ratio of cross-section after the generation cuts to the inclusive one. The errors associated with inclusive cross-sections are obtained from \texttt{MadGraph5}. The generation cuts are as follows: $p_{\mathrm{T},j}>5~\text{GeV}$, $p_{\mathrm{T},\gamma}>160~\text{GeV}$, $|\eta_{j/\gamma}|<3.0$, $\Delta R_{jj}>0.01$, $\Delta R_{j\gamma}>1.0$. Here, $p$ and $j$ both contain gluon, light quarks and $b$-quark.}
\label{tab:cs_cuts}
\end{table}

\subsection{Object definitions and event selection}
\label{sec:event_sel}

\begin{figure}[!tb]
\begin{center}
\subfloat[]{\label{fig:multijet}
\includegraphics[width=.45\textwidth]{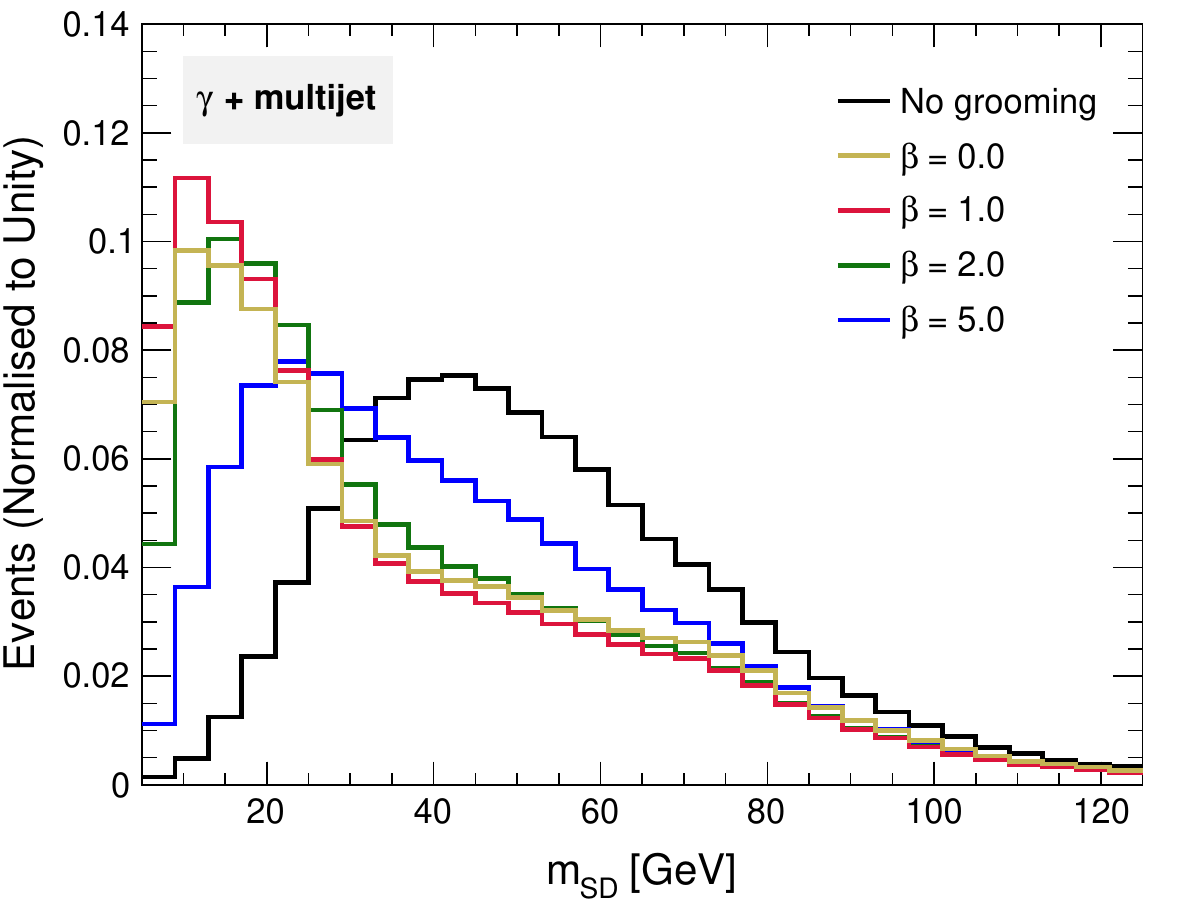}
}\qquad
\subfloat[]{\label{fig:alp}
\includegraphics[width=.45\textwidth]{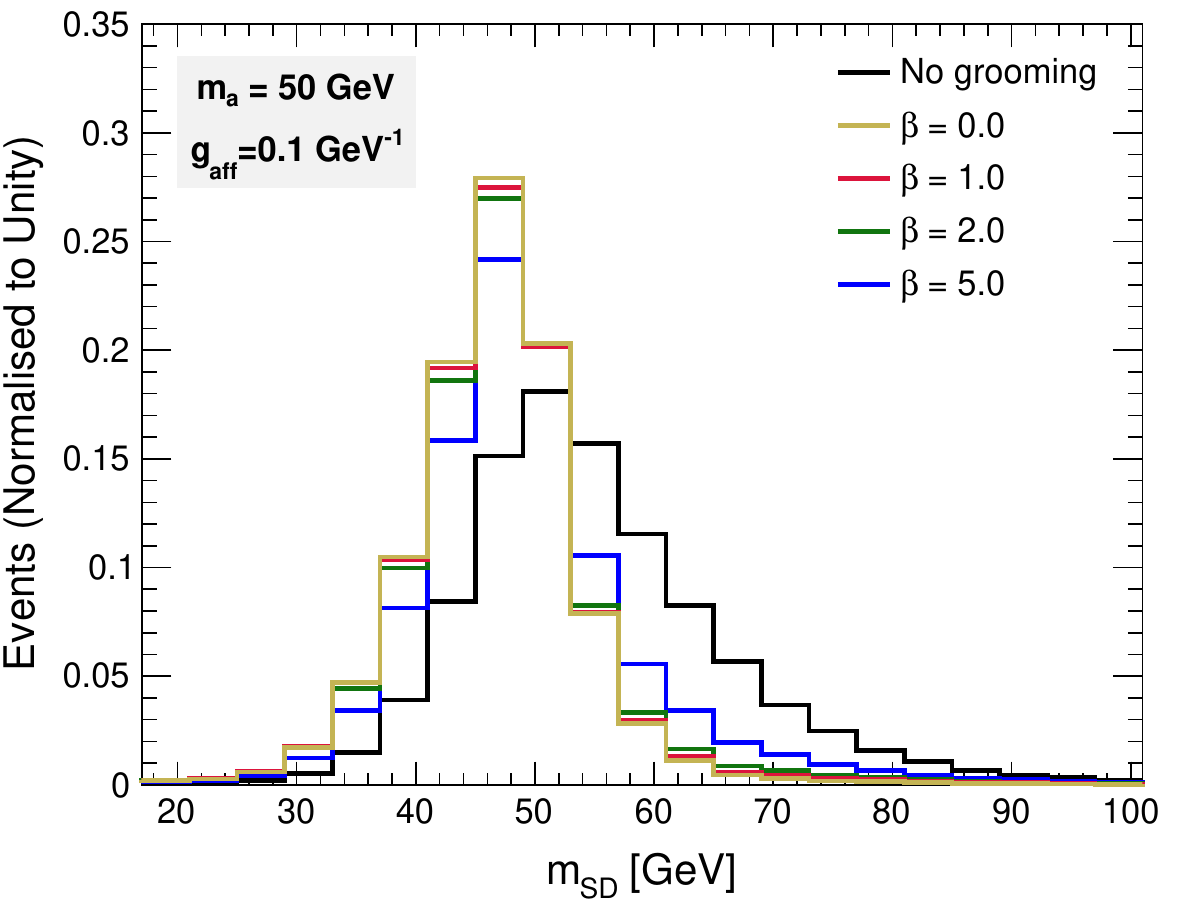}
}
\end{center}
\caption{The soft drop mass distributions of AK1.0 jet by changing the soft drop parameter $\beta$, with fixed $z_{cut}=0.1$, (a) for the multijet background, and (b) ALP signal of mass $m_a=50$ GeV.}
\label{fig:mSD_beta}
\end{figure}

At selection level, a large radius jet for full containment of the hadronisation products, coming from the final state $b$-quarks, is required. The reconstruction of jets is performed within the {\tt FastJet} framework~\cite{Cacciari:2011ma}, utilizing the AK1.0 anti-$k_{\textrm{T}}$ clustering algorithm, using the particle-flow objects in Delphes as input. 
The photon, used to trigger on, is required to have $p_{\mathrm{T},\gamma}>160$ GeV and $|\eta_\gamma|<2.1$~\cite{ATLAS:2018hbc,CMS:2019xai}. 
Further, a separation between the photon and AK1.0 jet is required to be $\Delta R > 2.2$~\footnote{The angular distance in the pseudorapidity ($\eta$) and azimuthal angle ($\phi$) plane is measured in units of $\Delta R =\sqrt{(\Delta\eta)^2 + (\Delta\phi)^2}$. Here, $\eta$ is defined in terms of polar angle ($\theta$) as $\eta=-{\rm ln~tan}(\theta/2)$.}~\cite{CMS:2019xai}. We apply the soft drop mass algorithm~\cite{Larkoski:2014wba} to groom the AK1.0 jet, using the parameters $z_{cut}$ = 0.1 and $\beta$ = 1~\cite{ATL-PHYS-PUB-2021-028} where, $z_{cut}$ denotes the energy threshold required to satisfy the soft drop condition during the declustering procedure. The soft drop condition is defined as $\frac{\text{min}(p_{\mathrm{T},1},p_{\mathrm{T},2})}{p_{\mathrm{T},1}+p_{\mathrm{T},2}}>z_{cut}~(\frac{\Delta R_{12}}{R_0})^\beta$. Here, $p_{\mathrm{T},i}$ are jet constituents' transverse momentum, $\Delta R_{12}$ is the angular separation between those constituents and $R_0$ is the jet radius parameter. 
The angular exponent $\beta \neq 0$ imposes an additional angular requirement on the soft drop condition. 
Specifically, the soft drop removes soft wide angle radiation from a jet, while keeping a fraction of the soft collinear radiation as long as $\beta>0$. 
Although the $\beta=0$ choice is widely used~\cite{CMS:2019xai}, this choice significantly reduces the signal efficiency the for lowest ALP mass ($m_a$ = 12 GeV) considered in this analysis while $\beta=1$ is found to be optimal across the entire mass range.
Fig.~\ref{fig:mSD_beta} represents the distribution of $\mathrm{m}_{\mathrm{SD}}$,  the invariant mass of the soft dropped AK1.0 jet. 
As shown, by removing the soft radiation, the grooming technique soften the $\mathrm{m}_{\mathrm{SD}}$ spectrum of multijet events, while it leaves the signal distribution almost unaltered. 
This, in turn, presents the advantage of allowing the background shape to be fitted using an exponentially falling distribution, which can be constrained by data on both sides of the resonant mass values (side-band method~\cite{ATL-PHYS-PUB-2020-028}). This is shown in Fig.~\ref{fig:mSD_beta} (a) for the case $\beta=1.0$, where the $\mathrm{m}_{\mathrm{SD}}$ distribution for multijet events presents an exponentially falling behaviour starting from values as low as 10 GeV.
In this particular case, a deviation can be seen starting from $\mathrm{m}_{\mathrm{SD}}=80$ GeV, consistent with resonant W- and Z-boson production.

In case of signal events, the two $b$-quarks from the ALP decay originate a clear 2-prong substructure signature that can be reconstructed within the AK1.0 jet. 
However, that is not the case for multijet background, where the final state features collimated radiation with merged hadronisation objects. 
The designed decorrelated tagger (DDT)~\cite{Dolen:2016kst,CMS:2017dcz} observable $\mathrm{N}_2^{\mathrm{DDT}}$, defined as
\begin{equation}
\mathrm{N}_2^{\mathrm{DDT}}(\rho,p_{\rm T}) \equiv \mathrm{N}_2 - 
\mathrm{N}_2^{10\%}(\rho,p_{\rm T}) \, ,
\label{eq:n2ddt}
\end{equation}
is used in the following, where $\mathrm{N}_2$ is a 2-prong discriminating observable and $\mathrm{N}_2^{10\%}$ is the value of $\mathrm{N}_2$ that rejects $90\%$ of the multijet background events, which depends on the jet $p_{\rm T}$ and the scaling variable $\rho$. 
The observable, $\mathrm{N}_2$ is constructed as the ratio of generalised energy correlation functions (ECF)~\cite{Moult:2016cvt} as 
\begin{equation}
\mathrm{N}_2 =\frac{_{2}e_{3}}{(_{1}e_{2})^2}\,,
\label{eq:N2}
\end{equation}
where the two point and three point ECFs, $_{1}e_{2}$ and $_{2}e_{3}$ are defined as 
\begin{align}
\begin{split}
_{1}e_{2} &= \sum_{1\leq i<j\leq N} z_{i}z_{j} \, \Delta R_{ij} \,, \\[1em]
_{2}e_{3} &= \sum_{1\leq i<j<k\leq N} z_{i}z_{j}z_{k} \min \left\{\Delta R_{ij} \Delta R_{ik}\,, \Delta R_{ij}  \Delta R_{jk}\,,     \Delta R_{ik} \Delta R_{jk}    \right\} \,. 
\label{eq:ECF}
\end{split}
\end{align}
Here, $z_i$ is the energy fraction of $i$th jet constituent, $z_i\equiv \frac{p_{\mathrm{T},i}}{p_{\mathrm{T},\mathrm{jet}}}$ and $\Delta R_{ij}$ correspond to the separation between the $i$th and $j$th jet constituents in the $\eta-\phi$ plane.
In general, $_{v}e_{n}$ depends on the $n$ energy fractions of the $N$ constituent particles inside the jet ($n\leq N$), relative to the jet $p_{\rm T}$ and the minimal product of $v$ number of pairwise angles between the $n$ selected particles. 
In a scenario where there is a perfect two-prong structure inside the jet, $(_{1}e_{2})^2\gg~_{2}e_{3}$~\cite{Moult:2016cvt}, therefore the smaller value of $\mathrm{N}_2$ in signal events can be used as discriminant against multijet background. 
At analysis level, in order to reduce the variable's dependency on the jet mass and transverse momentum, the $\mathrm{N}_2$ variable is estimated on the groomed AK1.0 jet.

\begin{figure}[!tb]
\centering
\includegraphics[width=.70\textwidth]{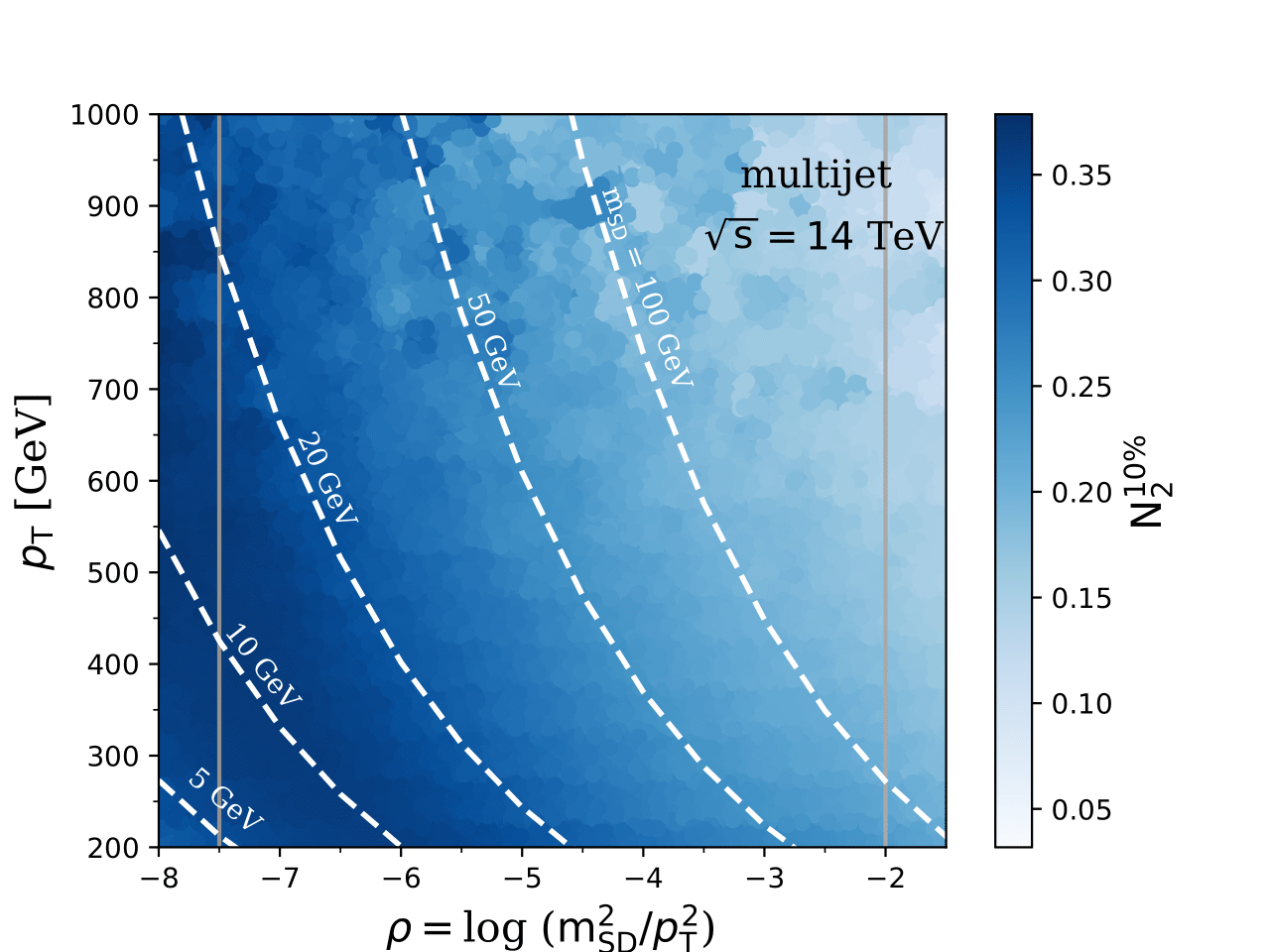}
\caption{The $10\%$ quantile of the $\mathrm{N}_2$ distribution in simulated multijet events, $\mathrm{N}_2^{10\%}$, used to define the $\mathrm{N}_2^{\mathrm{DDT}}$ variable. The distribution is shown in the $\rho-p_{\rm T}$ plane. The white dashed lines correspond to constant values of the jet mass, $\mathrm{m}_{\mathrm{SD}}$. We select the region $-7.5<\rho<-2.0$ for our analysis, as shown with vertical gray lines.}
\label{fig:N10}
\end{figure}

The dimensionless scaling variable $\rho$ defined as $\rho=\mathrm{log}(\mathrm{m}_{\mathrm{SD}}^2/p_{\rm T}^2)$~\cite{Dolen:2016kst} is designed to reject events having non-perturbative effects in the QCD jet mass spectrum. 
The selected $\rho$ region in the analysis is $-7.5\leq \rho\leq -2.0$, which is efficient for the signal process, and it corresponds to our region of interest, with $\mathrm{m}_{\mathrm{SD}}$ from 5 GeV to 100 GeV, for AK1.0 jet with $p_{\rm T}$ = 200 GeV. $\mathrm{N}_2^{10\%}$, shown in Fig.~\ref{fig:N10}, is evaluated in multiple bins of the $\rho-p_{\rm T}$ plane. Smoothing, using the distance-weighted $k$-nearest-neighbor (kNN) rule~\cite{5408784}, is applied to $\mathrm{N}_2^{10\%}$, to efficiently populate all the phase space. 
The signal efficiency of the $\mathrm{N}_2^{\mathrm{DDT}}<0$ requirement, varies roughly between $12\%$ and $26\%$, increasing the signal over background ratio, since by construction only 10\% of the background are retained, with the exception of $m_a=12$ GeV for which the efficiency drops to $\sim 1\%$. 
For the latter, given the high $p_{\rm T}$ of the AK1.0 jet, the $b$-quarks from the ALP decay are so  collimated that the two-prong structure is lost, resulting in a large loss in efficiency after the $\mathrm{N}_2^{\mathrm{DDT}}$ requirement. We will show later that this can be partly recovered if a lower threshold for the trigger photon can be achieved by the LHC experiments.   

\begin{figure}[!tb]
\centering
\includegraphics[width=.65\textwidth]{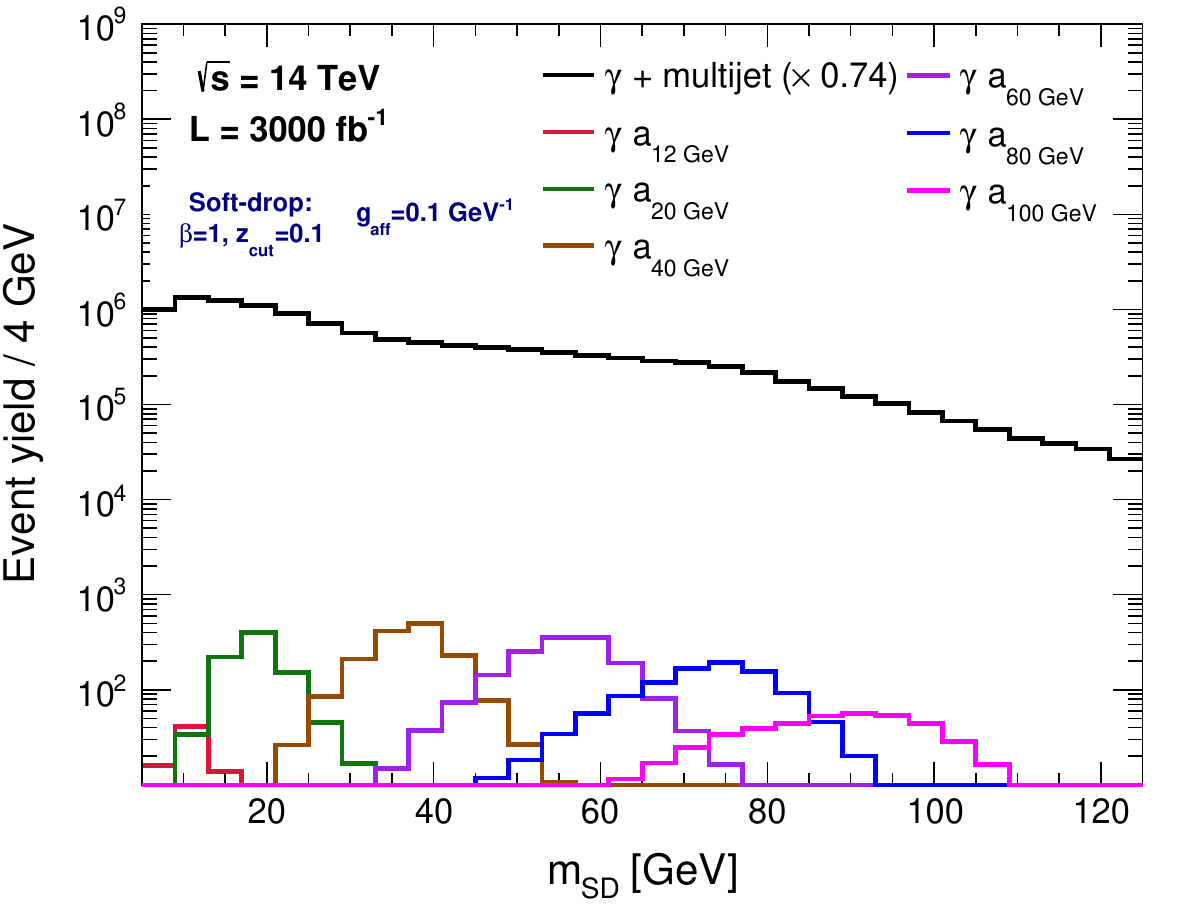}
\caption{The soft drop mass distributions of AK1.0 jet for the multijet background and ALP signal. The ALP signals are shown as $\gamma$ a$_{m_a}$, for example, $\gamma$ a$_{\rm 12~GeV}$ corresponds to signal process with $m_a=$ 12 GeV. The $\gamma+$multijet background includes contributions from $W+\gamma$ and $Z+\gamma$ processes. The event yield on the y-axis is evaluated after all the selection cuts (see Table~\ref{tab:cuts}). The background distribution is corrected with an overall factor of $0.74$~\cite{CMS:2017dcz}.}
\label{fig:mSD}
\end{figure}

After the requirements listed in Table~\ref{tab:cuts}, we show the mass distribution, $\mathrm{m}_{\mathrm{SD}}$, for the multijet background and ALP signal, normalized to $\mathcal{L}=3000$ fb$^{-1}$ of proton-proton collisions at center-of-mass energy $\sqrt{s}$ of 14 TeV, in Fig.~\ref{fig:mSD}. 
The shape of the $\mathrm{m}_{\mathrm{SD}}$ distribution for multijet events was found in good agreement with Ref.~\cite{CMS:2017dcz,CMS:2019xai}, correspondingly, we apply a data-Monte Carlo scale factor of $0.74$~\cite{CMS:2017dcz} to the multijet background cross-section as estimated by \texttt{Pythia8}, which after the generation requirements is 323 pb. The mass distributions of the ALP signal are shown for $g_{\mathrm{aff}}=0.1$ GeV$^{-1}$, for some benchmark masses of ALP. It can be noticed that as the generated ALP mass increases, the variance of the $\mathrm{m}_{\mathrm{SD}}$ resonance distribution increases as well. This can be understood as the cone size of the AK1.0 jet, with radius parameter $R=1.0$, starts to be insufficient to encompass all the hadronisation objects from a heavy ALP decay recoiling from a photon with $p_{\rm T}>160$ GeV, as the ALP becomes less boosted. 

\begin{table}[!tb]
    \centering
    \begin{tabular}{l | l}
    \toprule
    \multicolumn{2}{c}{Selection requirements} \tabularnewline \bottomrule
    $p_{\mathrm{T},\gamma}$          & $>160$ GeV\\
    $|\eta_\gamma|$                  & $<2.1$ \\
    $p_{\mathrm{T, AK1.0}}$          & $>200$ GeV\\
    $\Delta R(\gamma,~\text{AK1.0})$ & $>2.2$ \\
    $\rho$                           & $[-7.5,-2.0]$\\
    $\mathrm{N}_2^{\mathrm{DDT}}$    & $<0$\\ \toprule 
    \end{tabular}
    \caption{Summary of the selection requirements. The transverse momentum of the photon and AK1.0 jet is denoted as $p_{\mathrm{T},\gamma}$ and $p_{\mathrm{T, AK1.0}}$, respectively, and distance between them is referred as $\Delta R(\gamma,~\text{AK1.0})$. $\eta_\gamma$ is the pseudorapidity of the photon.  $\rho$ is a dimensionless scaling variable, $\rho=\mathrm{log}(\mathrm{m}_{\mathrm{SD}}^2/p_{\rm T}^2)$. The DDT observable is represented as $\mathrm{N}_2^{\mathrm{DDT}}$.}
    \label{tab:cuts}
\end{table}

\subsection{Flavour labelling and $b$-tagging on the simulated events}
\label{sec:di-btag}

Identifying hadronic jets containing $b$-hadrons ($b$-tagging), is an essential tool to improve signal sensitivity in searches where the final state contains $b$-quarks. 
The ATLAS and CMS experiments utilises highly performing Neural Network-based $b$-tagging algorithms~\cite{ATL-PHYS-PUB-2017-013,CMS:2017wtu} in many Higgs, Supersymmetry, top-quark analyses~\cite{ATLAS:2019bwq,ATLAS-CONF-2016-039,ATLAS:2017avc,CMS-PAS-SUS-10-011,CMS:2020zge,Khachatryan:2240708}.
The search for boosted Higgs production~\cite{ATL-PHYS-PUB-2023-021} at the LHC using GN2X  double $b$- and double $c$-tagging algorithms is taken as a reference for the ALP analysis proposed in this work.

\texttt{Delphes}, used for simulating the detector reconstruction, not having tracking reconstruction capability, uses a flavour dependent parametrized single jet $b$-tagging probability based on ATLAS and CMS published performance~\cite{ATL-PHYS-PUB-2015-022,CMS:2012feb}. 
The emulation of double $b$-tagging is developed and applied outside \texttt{Delphes} by the authors of this work.
In particular, we use a parametrisation of the boosted double $b$- and double $c$-tagging algorithm with a
working point of $80\%$ $bb$ tagging efficiency. The corresponding rejection efficiencies for multijet background are taken from the GN2X performance listed in Ref.~\cite{ATL-PHYS-PUB-2023-021}. The parametrisation involves identifying smaller radius jets, called subjets, inside the AK1.0 jet and matching them to a hadron in the event to identify  different double flavour jet categories, and it is explained in what follows. 

The multijet background events which passes the soft-drop grooming and $\mathrm{N}_2^{\mathrm{DDT}}<0$ requirements, discussed Section~\ref{sec:event_sel}, mostly contain AK1.0 jets presenting a two prong substructure. These may originate from the initial state partons in the following three ways: two quarks ($qq'$, $q/q'=u,d,c,s,b$), or a quark and a gluon ($qg$), or two gluons (gg), produced back-to-back to the high $p_{\rm T}$ photon.
In order to categorise the flavour content of these events, ATLAS uses variable radius (VR) anti-$k_{\rm T}$ jets~\cite{Krohn:2009zg,ATL-PHYS-PUB-2017-010}, since fixed radius jet algorithms\cite{Cacciari:2008gp,CATANI1993187,Wobisch:1998wt}  introduce overlapping subjets and reduce the efficiency for flavour labelling in case of low ALP mass where the decay products are highly collimated. 

\begin{figure}[!tb]
\begin{center}
\subfloat[]{\label{fig:12}
\includegraphics[width=.45\textwidth]{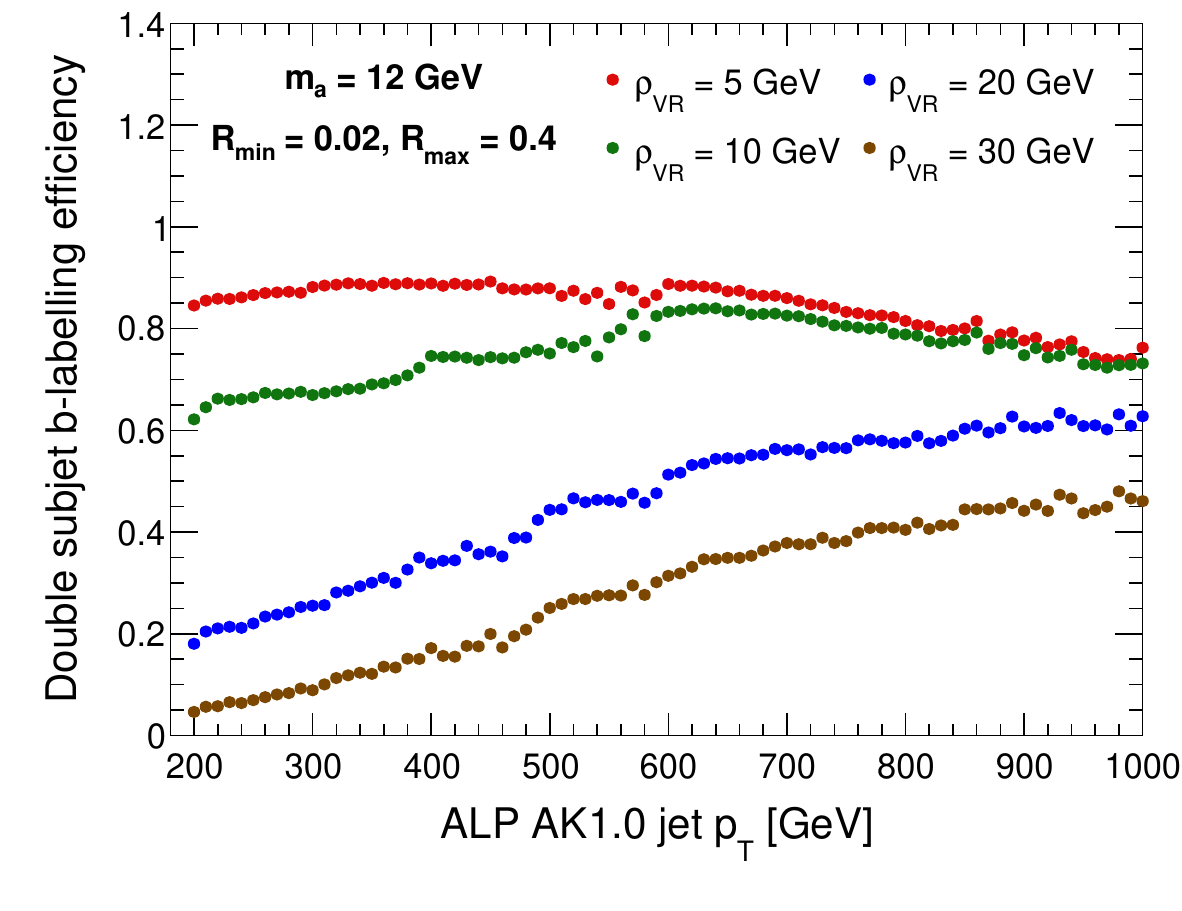}
}\qquad
\subfloat[]{\label{fig:40}
\includegraphics[width=.45\textwidth]{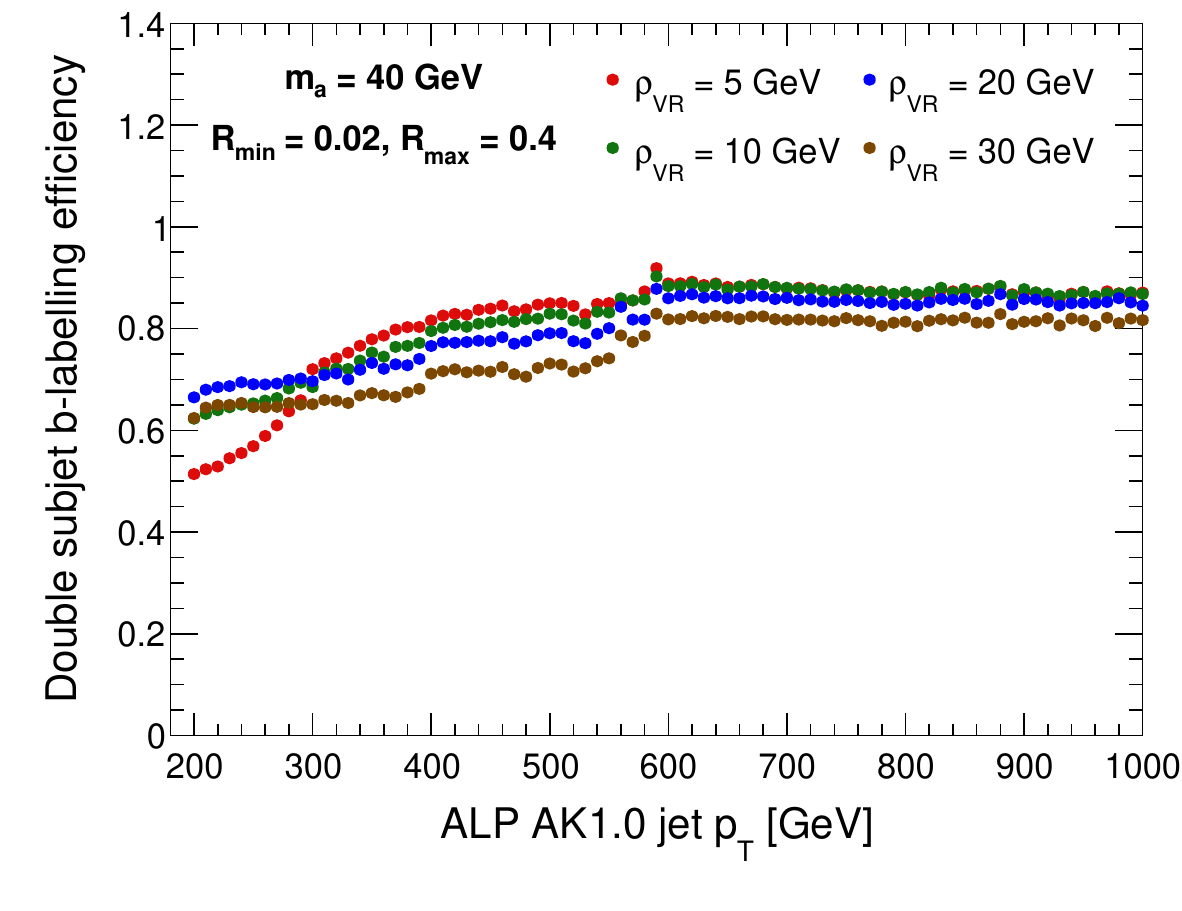}
}\qquad
\subfloat[]{\label{fig:60}
\includegraphics[width=.45\textwidth]{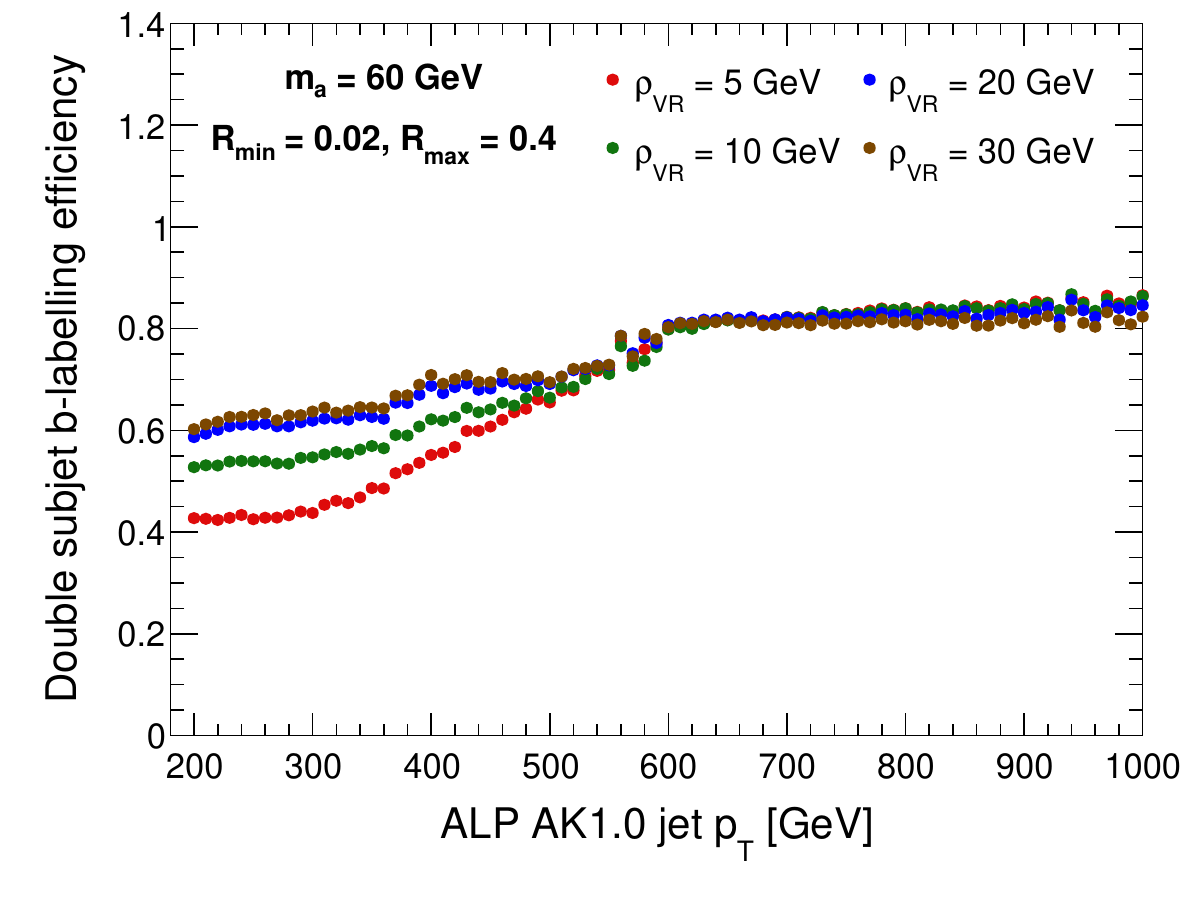}
}\qquad
\subfloat[]{\label{fig:100}
\includegraphics[width=.45\textwidth]{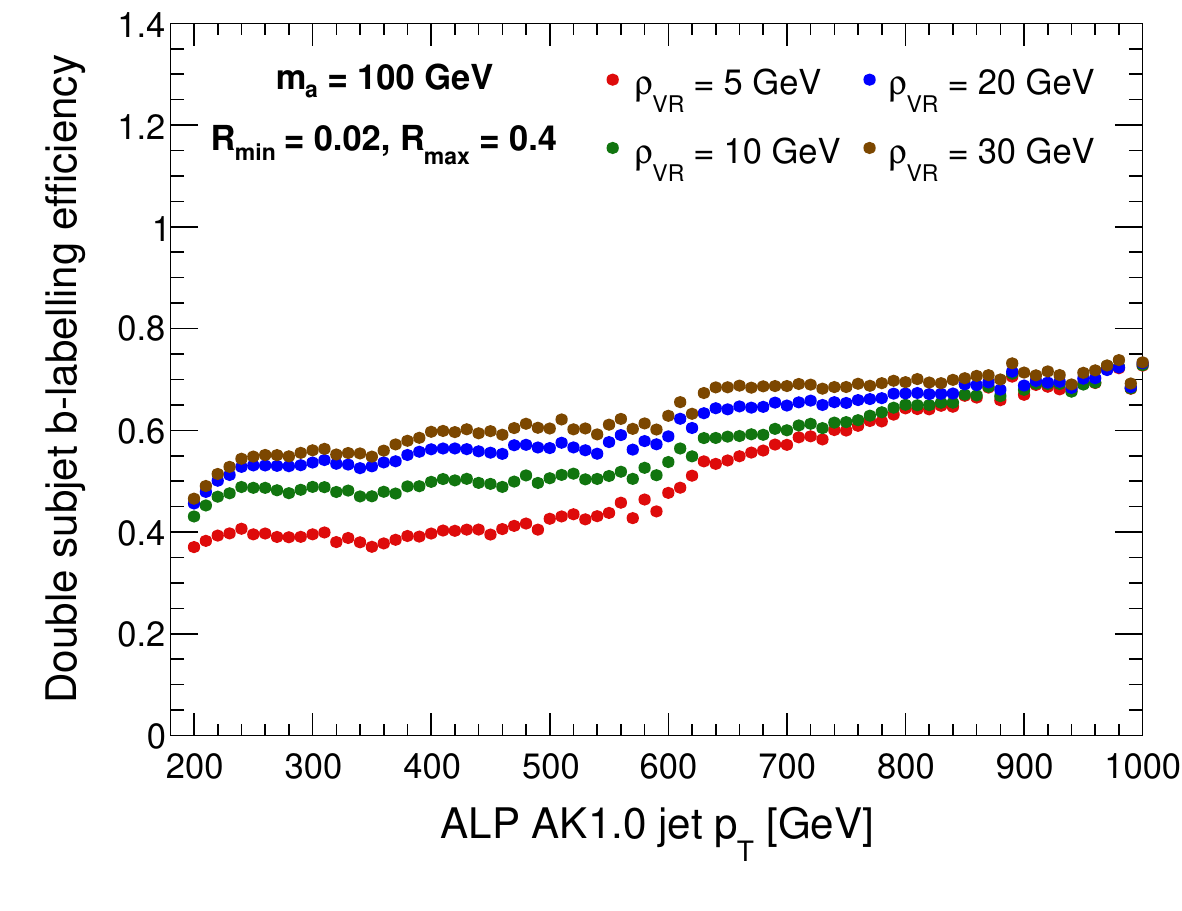}
}
\end{center}
\caption{Comparison of double subjet $b$-labelling efficiency for different choices of the $\rho$\mbox{\small $_{\mathrm{VR}}$} parameter while constructing the VR subjets. The efficiencies are shown for the simulated ALP signal with masses, (a) $m_a=$ 12 GeV, (b) $m_a=$ 40 GeV (c) $m_a=$ 60 GeV and (d) $m_a=$ 100 GeV. The $R_{\mathrm{min}}$ and $R_{\mathrm{max}}$ values are set to 0.02 and 0.4~\cite{ATL-PHYS-PUB-2017-010}, respectively.}
\label{fig:Rho_choice}
\end{figure}

The VR algorithm can differentiate the substructure coming from a hard parton than from soft radiation because the radius of the constructed jet changes according to the jet transverse momentum. Three parameters are required to define the VR subjets: $\rho$\mbox{\small $_{\mathrm{VR}}$}, $R_{\mathrm{min}}$ and $R_{\mathrm{max}}$. $\rho$\mbox{\small $_{\mathrm{VR}}$} is a dimensionful constant and the effective jet radius depends on the jet transverse momentum as, $R_{\mathrm{eff}}(p_{\rm T})=$$\rho$\mbox{\small $_{\mathrm{VR}}$}$/p_{\rm T}$. The parameters, $R_{\mathrm{min}}$ and $R_{\mathrm{max}}$, ensure that the effective jet radius do not fall below the detector resolution at high $p_{\rm T}$ and jets do not become arbitrarily large at small $p_{\rm T}$, respectively. These three parameters need to be optimised to improve signal acceptance or background rejection for specific scenarios. The minimum and maximum values for the radius parameter are optimised in Ref.~\cite{ATL-PHYS-PUB-2017-010} and we use them in our analysis, $R_{\mathrm{min}}$ = 0.02 and $R_{\mathrm{max}}$ = 0.4. However, we optimise $\rho$\mbox{\small $_{\mathrm{VR}}$} in the following paragraph since a wide range of masses for the ALP signal are considered in this work.

We choose $\rho$\mbox{\small $_{\mathrm{VR}}$} to maximise the double subjet $b$-labelling efficiency, as a function of the transverse momentum of the AK1.0 jet. The efficiency is defined as the ratio between the number of AK1.0 jets having two $b$-subjets with $p_{\mathrm{T},\text{subjet}}>7$ GeV and the total number of AK1.0 jets. The flavour matching procedure is discussed in the following paragraph~\footnote{Note that the same flavour matching method is used first to set the optimal $\rho$\mbox{\small $_{\mathrm{VR}}$} parameter, and later to categorise the AK1.0 jet into different flavour composition.}. 
A comparison of these efficiencies for different choices of the $\rho$\mbox{\small $_{\mathrm{VR}}$} parameter is shown in Fig.~\ref{fig:Rho_choice} for ALP masses of 12 GeV, 40 GeV, 60 GeV and 100 GeV. A smaller $\rho$\mbox{\small $_{\mathrm{VR}}$} value is preferred for a light ALP, while the efficiency is higher with larger $\rho$\mbox{\small $_{\mathrm{VR}}$} parameter for a heavier ALP. These preferred choices for the $\rho$\mbox{\small $_{\mathrm{VR}}$} value, specifically $\rho$\mbox{\small $_{\mathrm{VR}}$}$=5$ GeV for $m_a=12$ GeV, 20 GeV and 30 GeV; $\rho$\mbox{\small $_{\mathrm{VR}}$}$=10$ GeV for $m_a=40$ GeV and 50 GeV; $\rho$\mbox{\small $_{\mathrm{VR}}$}$=20$ GeV for $m_a=60$ GeV and 70 GeV; $\rho$\mbox{\small $_{\mathrm{VR}}$}$=30$ GeV for $m_a=80$ GeV, 90 GeV and 100 GeV; are used while constructing the two subjets with the VR algorithm, for both the signal and multijet background.

Next, we match each VR subjet to a hadron. To be compliant with the labelling procedure used to derive the parametrized efficiency in Ref.~\cite{ATL-PHYS-PUB-2020-019}, we assign a $b$-, $c$- and light flavor, to each VR subjet. We first identify $b$-hadrons and $c$-hadrons in the event from the list of all stable particles provided by \texttt{Pythia8} using their PDG codes. The hadrons must have $p_{\mathrm{T},\text{hadron}}>5$ GeV, and are required to be within $\Delta R<1.0$ of the AK1.0 jet axis. 
Events are selected if they contain at least two VR subjets with $p_{\mathrm{T},\text{subjet}}>7$ GeV, similar to earlier double subjet $b$-labelling efficiency, and the subjets fall inside the AK1.0 jet. Thus the selected events have at least two VR subjets and identified hadrons within $\Delta R<1.0$ of the AK1.0 jet. The VR subjet is called $b$-jet (\textbf{b}) when a $b$-hadron is found within $\Delta R<0.3$~\cite{ATL-PHYS-PUB-2020-019} of the VR subjet. If there is no nearby $b$-hadron but a $c$-hadron is found with $\Delta R(\text{$c$-hadron, VR-subjet})<0.3$, the VR subjet is labelled as $c$-jet (\textbf{c}). In case the $b$-hadron or $c$-hadron is within $\Delta R<0.3$ of multiple VR subjets, the hadron is associated to the closest VR subjet. Otherwise, the VR subjet is labelled as a light jet (\textbf{l}). We summmarise all the above discussion in Table~\ref{tab:matching}. 

\begin{table}[tb]
\centering
\begin{tabular}{l|l}\toprule
\multicolumn{2}{c}{VR parameters} \\\hline
\multicolumn{1}{l|}{$p_{\mathrm{T},\text{subjet}}$ } & $>7$ GeV\\
\multicolumn{1}{l|}{$R_{\mathrm{min}}$} & 0.02\\
\multicolumn{1}{l|}{$R_{\mathrm{max}}$} & 0.4\\\hline
  \multicolumn{1}{l|}{$m_a$ [GeV]}  & $\rho$\mbox{\small $_{\mathrm{VR}}$} [GeV]\\\cline{1-2}
  12, 20, 30 & 5\\
  40, 50 & 10\\
  60, 70 & 20\\
  80, 90, 100 & 30\\\toprule\toprule
\multicolumn{2}{c}{Matching parameters} \\\hline
\multicolumn{1}{l|}{$p_{\mathrm{T},\text{hadron}}$ } & $>5$ GeV\\  
\multicolumn{1}{l|}{$\Delta R$(hadron, VR)} & $<0.3$\\
\multicolumn{1}{l|}{$\Delta R$(hadron/VR,~AK1.0)} & $<1.0$\\\toprule
\end{tabular}
\caption{Summary of the parameters used in variable radius (VR) subjet construction (top panel) and the matching procedure (bottom panel). The radius parameter for the VR subjet is $R_{\mathrm{eff}}(p_{\rm T})=$$\rho$\mbox{\small $_{\mathrm{VR}}$}$/p_{\rm T}$. The transverse momentum of the VR subjet is $p_{\mathrm{T},\text{subjet}}$. The minimum and maximum radius parameters in the VR algorithm are $R_{\mathrm{min}}$ and $R_{\mathrm{max}}$, respectively. If a $b$-hadron is found, the jet is called $b$-jet, otherwise if a $c$-hadron is found, the jet is called $c$-jet, if no $c$- or $b$-hadrons are present the jet is referred to as light jet. Here, jet refers to the VR subjets.}
\label{tab:matching}
\end{table}

Since we have at least two flavour matched VR subjets, the AK1.0 jets can be broken down to the following flavour compositions: $bb$, $bc$, $bl$, $cc$, $cl$, and $ll$. The signal events having at least two VR subjets inside the AK1.0 jet labelled as $b$-jets ($bb$) are approximately $71\%$ for a 12 GeV ALP signal, which increases to around $90\%$ for $m_a=20$ GeV and 30 GeV. The percentage of signal events passing that criteria decreases gradually afterwards, reducing down to roughly $62\%$ for $m_a=100$ GeV, since the jet radius becomes insufficient to include all hadronisation objects, particularly for heavier ALP (see Fig.~\ref{fig:Rho_choice}). In the case of the multijet sample, around $85\%$ of the AK1.0 jets have no $b$- or $c$-subjets. The fraction of AK1.0 jets with subjet flavour in the various categories is tabulated in Table~\ref{tab:flav_comb}, for $\rho$\mbox{\small $_{\mathrm{VR}}$}$=30$ GeV. The fractional values hardly change with $\rho$\mbox{\small $_{\mathrm{VR}}$} parameter.

\begin{table}[tb]
    \centering
    \begin{tabular}{l|c|c|c|c|c|c}
    \toprule
    \multicolumn{7}{c}{Multijet background}\\\hline
    Category & $bb$ & $bc$ & $bl$ & $ll$ & $cl$ & $cc$  \\ 
    Fractional composition ($\%$) & $0.6$ & $0.5$ & $1.2$ & $84.4$ & $10.5$ & $2.8$\\ 
    GN2X efficiency~\cite{ATL-PHYS-PUB-2023-021} & $0.59$ & $0.21$ & $0.07$ & $0.21\times 10^{-3}$ & $0.41\times 10^{-2}$ & $0.02$ \\\toprule
    \end{tabular}
    \caption{The fraction of AK1.0 jets in the multijet sample, sorted by flavour composition, that have at least two VR subjets. The values are shown for $\rho$\mbox{\small $_{\mathrm{VR}}$}$=30$ GeV and remain roughly the same with different choices of the $\rho$\mbox{\small $_{\mathrm{VR}}$} parameter. The GN2X efficiency for background in each category are shown for the working point of $80\%$ $bb$ tagging efficiency, from Ref.~\cite{ATL-PHYS-PUB-2023-021}.}
    \label{tab:flav_comb}
\end{table}

\subsection{Signal extraction and background estimation}
\label{sec:Nevents}

\begin{figure}[!tb]
\centering
\includegraphics[width=.65\textwidth]{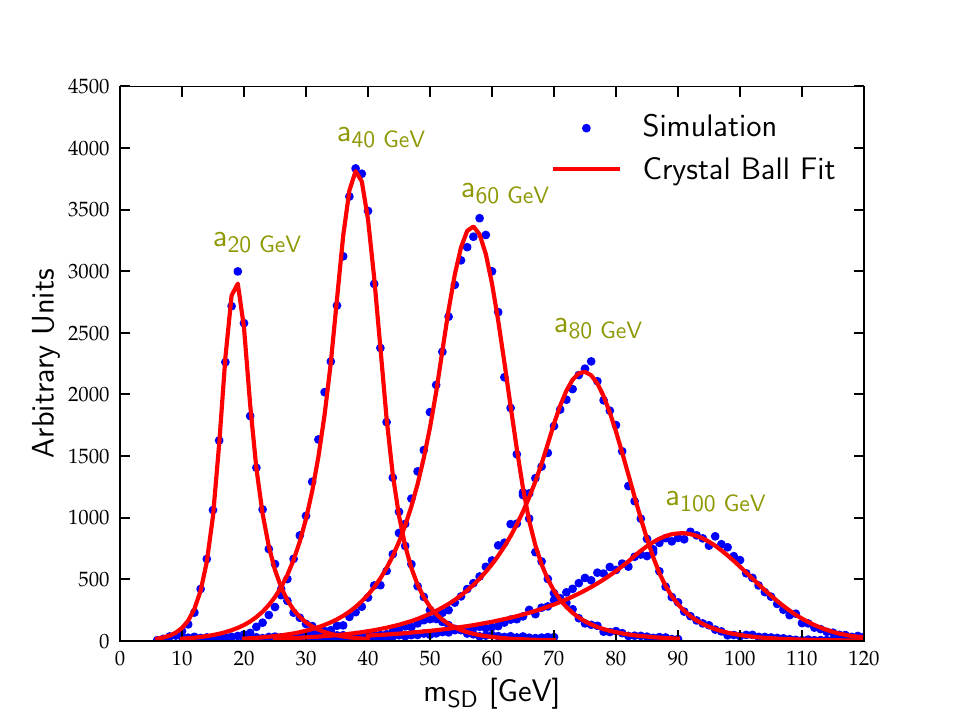}
\caption{The $\mathrm{m}_{\mathrm{SD}}$ distribution for the simulated signal samples and the Crystal Ball fits, in blue and red color, respectively. The fits are shown for signal events with $m_a=$ 20, 40, 60, 80 and 100 GeV, after all the requirements mentioned in Table~\ref{tab:cuts}.}
\label{fig:fit}
\end{figure}

In light of the precise background estimation provided by the $\mathrm{m}_{\mathrm{SD}}$ side-band fit, the final result is derived by counting the number of signal and background events, after the selection requirements listed in Table~\ref{tab:cuts}, in a mass window centered  around the ALP reconstructed $\mathrm{m}_{\mathrm{SD}}$.  While complex analysis strategies involving substructure observables could increase the sensitivity of the resonance search and should be attempted in the experimental analysis, these are not considered in this work since such approaches would require several steps that use collider data for validation. 
For each mass point, we estimate the $\mathrm{m}_{\mathrm{SD}}$ resolution using a double-sided Crystal Ball function~\cite{Oreglia:1980cs,ATLAS-CONF-2014-031}, which is found to be robust against the change in shape of the resonance peak for different ALP masses. 
The Crystal Ball function, consists of a Gaussian function to fit the core of the distribution in combination of a power law function to fit both the low and high $\mathrm{m}_{\mathrm{SD}}$ regions. 
The Crystal Ball function is defined as:
 \begin{equation}
 \begin{split}
   & f(\mathrm{m}_{\mathrm{SD}}|N,\mu,\sigma,\alpha_{\mathrm{low}},\alpha_{\mathrm{high}},n_{\mathrm{low}},n_{\mathrm{high}}) = \\ \\
   & N\times 
    \begin{cases}
    \mathrm{e}^{-0.5(\frac{\mathrm{m}_{\mathrm{SD}}-\mu}{\sigma})^2}, & \mbox{if $-\alpha_{\mathrm{low}} \leq \frac{\mathrm{m}_{\mathrm{SD}}-\mu}{\sigma} \leq \alpha_{\mathrm{high}}$~,}\\
    \mathrm{e}^{-0.5\alpha_{\mathrm{low}}^{2} } \left[\frac{\alpha_{\mathrm{low}}}{n_{\mathrm{low}}} \left(\frac{n_{\mathrm{low}}}{\alpha_{\mathrm{low}}} - \alpha_{\mathrm{low}} -\frac{\mathrm{m}_{\mathrm{SD}}-\mu}{\sigma} \right)\right]^{-n_{\mathrm{low}}},  & \mbox {if $\frac{\mathrm{m}_{\mathrm{SD}}-\mu}{\sigma} < -\alpha_{\mathrm{low}}$~,}\\
    \mathrm{e}^{-0.5\alpha_{\mathrm{high}}^{2} } \left[\frac{\alpha_{\mathrm{high}}}{n_{\mathrm{high}}} \left(\frac{n_{\mathrm{high}}}{\alpha_{\mathrm{high}}} - \alpha_{\mathrm{high}}  + \frac{\mathrm{m}_{\mathrm{SD}}-\mu}{\sigma} \right)\right]^{-n_{\mathrm{high}}},  & \mbox {if $\frac{\mathrm{m}_{\mathrm{SD}}-\mu}{\sigma} > \alpha_{\mathrm{high}}$~,}\\
    \end{cases}
    \label{eq:CBF}
\end{split}
 \end{equation}
where $N$ is an overall normalisation parameter. The Gaussian part of the Crystal Ball distribution has mean value $\mu$ and width $\sigma$, while $\alpha_{\mathrm{low}}$ and $\alpha_{\mathrm{high}}$ parameterise the points where the power law function takes over. 
The corresponding exponents of the power law function are denoted as $n_{\mathrm{low}}$ and $n_{\mathrm{high}}$, respectively. 
In Fig.~\ref{fig:fit}, we show the $\mathrm{m}_{\mathrm{SD}}$ distribution for the simulated signal samples,  together with their corresponding Crystal Ball fit, for ALP masses of 20 GeV, 40 GeV, 60 GeV, 80 GeV and 100 GeV. 

\begin{table}[tb!]
\center
\begin{tabular}{c|c|c|c|c|c}
\bottomrule
\multirow{3}{*}{ $m_a$ [GeV] }  & \multicolumn{2}{c|}{ Fitted parameters [GeV]}  &  \multirow{3}{*}{ $\epsilon$ ($\times 10^{-2}$) } & \multicolumn{2}{c}{Yield at $\mathcal{L}=3000$ fb$^{-1}$} \\\cline{2-3}\cline{5-6}

& \multirow{2}{*}{$\mu$ } & \multirow{2}{*}{$\sigma$} && \multirow{2}{*}{\makecell{$S$ ($g_{\mathrm{aff}}=0.1$ GeV$^{-1}$})} &  \multirow{2}{*}{$N_{\rm B}$ }\\ 
&& &&&\\ \hline

 $12$  & $10.91\pm 0.16$ & $2.21\pm 0.17$   & $0.10$ & 30  & $4040$ \\
 $20$  & $18.70\pm 0.03$ & $2.46\pm 0.05$   & $1.00$ & 400 & $7820$ \\
 $30$  & $28.22\pm 0.03$ & $3.39\pm 0.04$   & $1.50$ & 590 & $7550$ \\
 $40$  & $38.16\pm 0.04$ & $3.95\pm 0.07$   & $1.90$ & 660 & $7810$ \\
 $50$  & $47.68\pm 0.03$ & $4.68\pm 0.05$   & $2.20$ & 710 & $11000$ \\
 $60$  & $56.86\pm 0.06$ & $5.76\pm 0.08$   & $2.40$ & 670 & $11310$ \\
 $70$  & $66.19\pm 0.06$ & $6.70\pm 0.08$   & $2.20$ & 550 & $11620$ \\
 $80$  & $74.80\pm 0.07$ & $7.33\pm 0.10$   & $1.80$ & 400 & $11770$ \\
 $90$  & $82.01\pm 0.13$ & $9.62\pm 0.14$   & $1.60$ & 310 & $11970$ \\
 $100$ & $90.55\pm 0.20$ & $10.96\pm 0.21$   & $1.10$ & 180 & $7310$  \\\toprule
\end{tabular}
\caption{The mean values, $\mu$ and widths, $\sigma$ obtained after fitting the ALP reconstructed $\mathrm{m}_{\mathrm{SD}}$ mass variable with the Crystal Ball function, along with the signal efficiency, $\epsilon$. The signal yields, $S$ for $g_{\mathrm{aff}}=0.1$ GeV$^{-1}$ and background yields, $N_{\rm B}$, are also shown at integrated luminosity of $3000$ fb$^{-1}$, for the simulated MC samples.}
\label{tab:final_yield}
\end{table}

The mass window, defined as the interval $[\mu - \sigma, \mu + \sigma]$, is used for the search of each ALP mass hypothesis and is tabulated in Table~\ref{tab:final_yield}, together with the corresponding signal efficiency and multijet background yield for proton-proton collisions at center-of-mass energy $\sqrt{s}=14$ TeV and 3000 fb$^{-1}$ of integrated luminosity. 
In this work, the multijet background is estimated with the generators chosen in line with what is used by LHC collaborations~\cite{ATL-PHYS-PUB-2021-028,ATLAS:2018hbc,CMS:2019xai,CMS:2017dcz}. The real experimental analysis will have to implement data-driven background techniques, which on one side will render the analysis insensitive to modeling uncertainties, and on the other side will allow the reduction of statistical uncertainties. This is due to the large sample of QCD events recorded by the collaboration, whose size is hard to match by simulations, given the amount of resources necessary to generate a comparable number of multijet events. 

As discussed earlier, the low signal efficiency at $m_a=12$ GeV results from the lack of 2-prong substructure at high jet $p_{\rm T}$ and therefore a low $\mathrm{N}_2^{\mathrm{DDT}}$ efficiency. 
To be noticed that since for high values of the ALP mass, the decay products tend to be less collimated, and therefore less likely to be contained in the cone of the AK1.0 jet, the width of the $\mathrm{m}_{\mathrm{SD}}$ distribution increases.

\section{Results}
\label{sec:results}

In this section, we present our final results in terms of $95\%$ confidence level (CL) upper exclusion limits on the $b\bar{b}\gamma$ production cross-section and the ALP-fermion coupling, $g_{\mathrm{aff}}$. The exclusion limits are evaluated in Section~\ref{sec:UL}. We further discuss how to improve the sensitivity by lowering the photon $p_{\rm T}$ threshold in Section~\ref{sec:low_pho_pt}. The results are complemented with astrophysical and collider bounds in Section~\ref{sec:summ_result}.

\subsection{Projected sensitivities on the cross-section and coupling}
\label{sec:UL}

\begin{figure}[!tb]
\begin{center}
\subfloat[]{\label{fig:multijet}
\includegraphics[width=.45\textwidth]{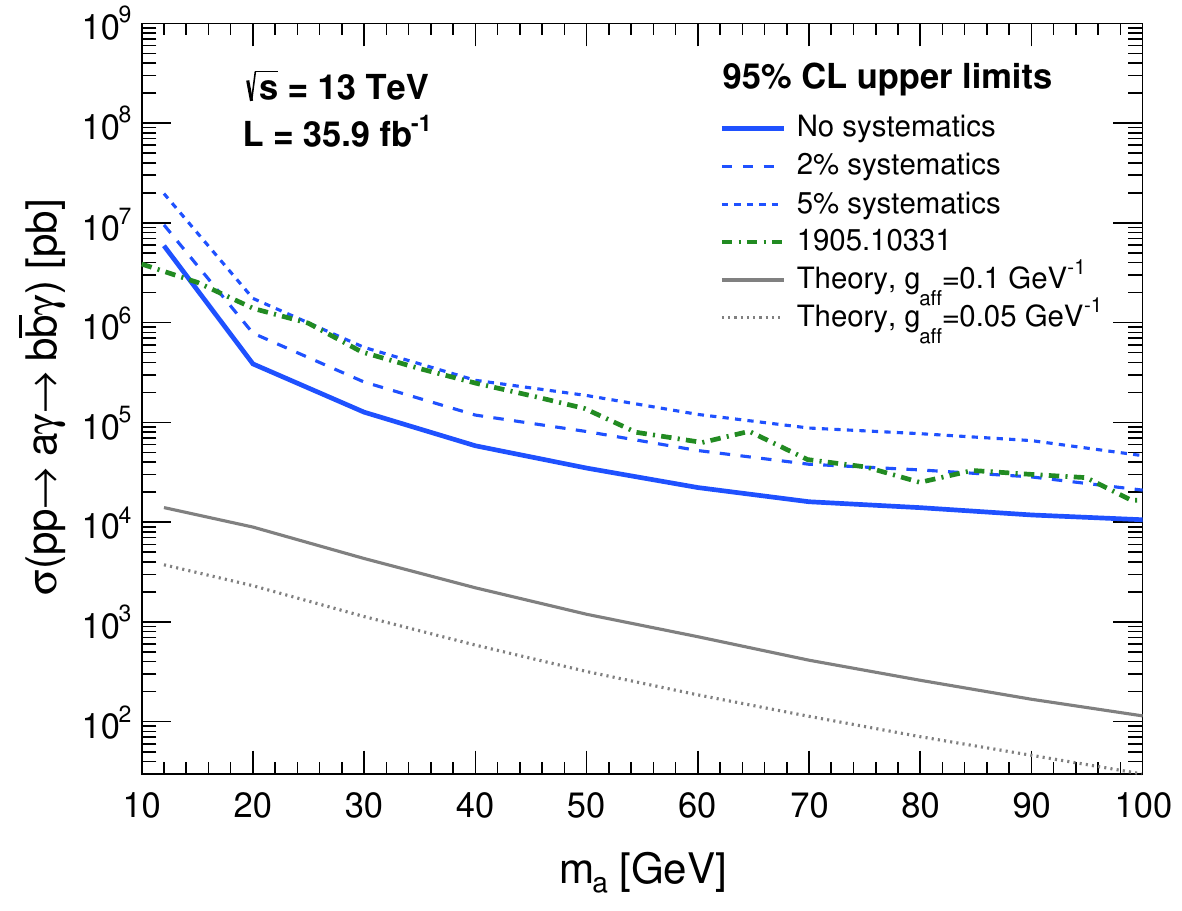}
}\qquad
\subfloat[]{\label{fig:alp}
\includegraphics[width=.45\textwidth]{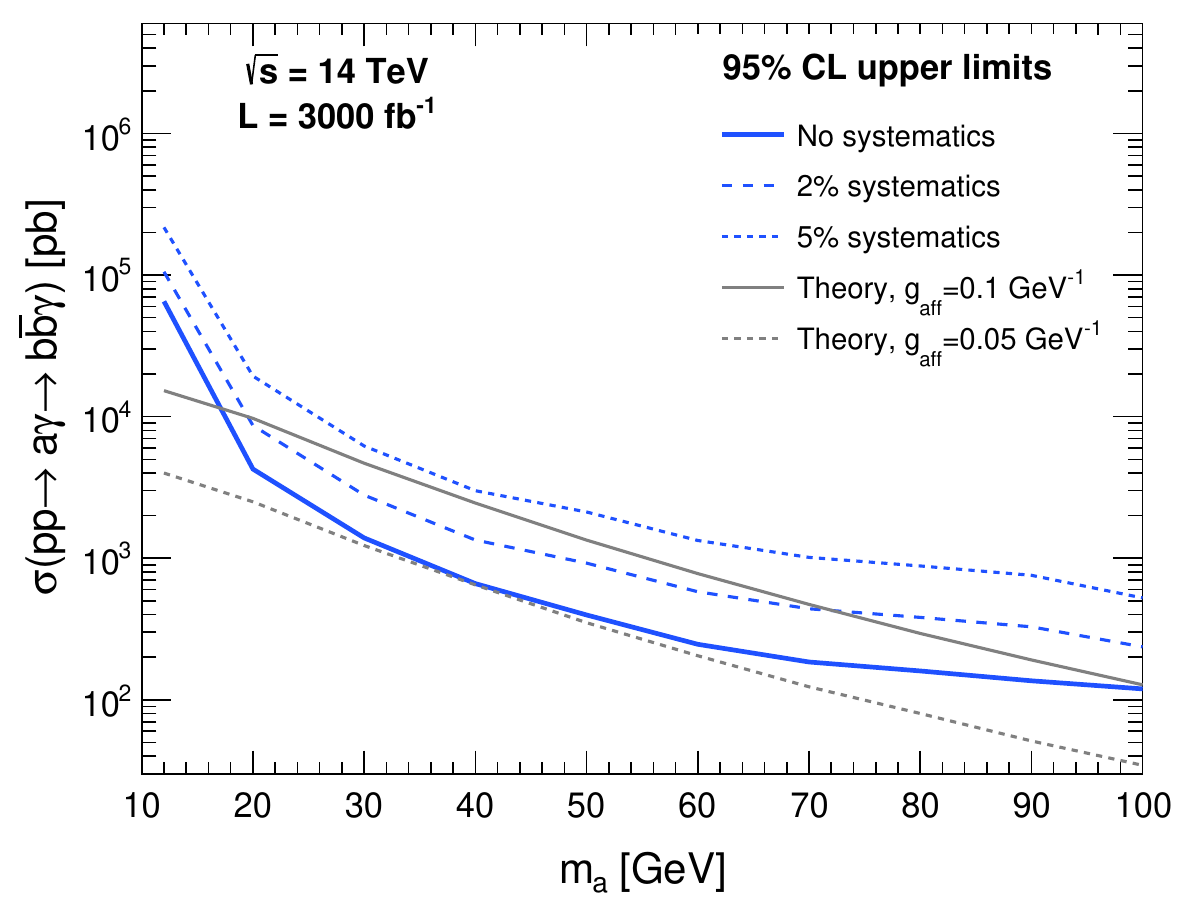}
}
\end{center}
\caption{Upper limit at $95\%$ C.L. on $\sigma(pp\to a\gamma\to b\bar{b}\gamma)$, as a function of the ALP mass. The limits are shown for (a) $\sqrt{s}=13$ TeV with $\mathcal{L}=35.9$ fb$^{-1}$ and (b) $\sqrt{s}=14$ TeV with $\mathcal{L}=3000$ fb$^{-1}$. The solid and (shorter) dashed blue lines correspond to neglecting systematic uncertainties or including them at the $2\%$ ($5\%$) level, respectively. The solid and dashed gray lines correspond to the theoretical prediction for the cross-section in the ALP model with $g_{\mathrm{aff}}=0.1$ GeV$^{-1}$ and 0.05 GeV$^{-1}$, respectively. 
For comparison, the dashed green line corresponds to the projected upper limit on the cross-section for an analogous process, $\sigma(pp\to Z'\gamma\to q\bar{q}\gamma)$ from Ref.~\cite{CMS:2019xai}.}
\label{fig:UL_sigma}
\end{figure}

The upper limit on the cross-section is evaluated with the signal significance formula~\cite{Cowan:2010js, Cowan:2012}, $S/\sqrt{N_{\rm B}}>N_{\mathrm{CL}}$, where $N_{\mathrm{CL}}=2$ corresponds to $95\%$ confidence interval. The signal yield, $S$ can be computed as: 
\begin{equation}
S = \sigma(pp\to a\gamma\to b\bar{b}\gamma)\cdot \epsilon\cdot \mathcal{L}~,
\nonumber
\end{equation} 
where $\sigma(pp\to a\gamma\to b\bar{b}\gamma)$ is the signal production cross-section, 
$\mathcal{L}$ is the integrated luminosity, and $\epsilon$ is the signal efficiency. 
After repeating the analysis for each $m_a$, we obtain $\epsilon (m_a)$ and $N_{\rm B} (m_a)$ as listed in Table~\ref{tab:final_yield}. 
In Fig.~\ref{fig:UL_sigma}, the $95\%$ CL upper limit on $\sigma(pp\to a\gamma\to b\bar{b}\gamma)$ is shown  as a function of ALP mass, by the solid blue line in the absence of systematic uncertainty, and by the (smaller) dashed blue line in the case of a $2\%$ ($5\%$) systematic uncertainty~\footnote{Adding $x\%$ systematic uncertainty, the signal significance formula is $S/\sqrt{N_{\rm B}+(x\times N_{\rm B}/100)^2}$.}. The size of the systematic uncertainties chosen are in line with current estimations by LHC experiments for jet mass scale and resolution~\cite{CMS:2019xai}. The limits are shown for  $\sqrt{s}=13$ TeV and $\mathcal{L}=35.9$ fb$^{-1}$, in the left panel of Fig.~\ref{fig:UL_sigma}, and for $\sqrt{s}=14$ TeV and $\mathcal{L}=3000$ fb$^{-1}$, in the right panel of Fig.~\ref{fig:UL_sigma}. In the latter case, the upper limits are stronger and vary between $6.5\cdot 10^4$ pb for an ALP mass of 12 GeV, and 119 pb for an ALP mass of 100 GeV. The limits weaken to $1.1\cdot 10^5$ pb ($2.2\cdot 10^5$ pb) and 237 pb (524 pb) upon adding $2\%$ ($5\%$) systematic uncertainty, respectively.

The limits are compared to the predicted cross-sections in the current ALP model, for different values of $g_{\mathrm{aff}}$, as shown by the gray lines in Fig.~\ref{fig:UL_sigma}. This shows that the HL-LHC will be sensitive to couplings in the range $g_{\rm aff}\sim (0.05-0.1)$ GeV$^{-1}$, for an ALP mass in the range $m_a \sim (20-100)$ GeV. The sensitivity around the lowest mass value considered in this work, \textit{i.e.} $m_a=12$ GeV, is significantly weaker. In the next subsection, we will discuss the possibility to improve this sensitivity with modified trigger requirements. We also compare our bounds with the upper limits presented in Ref.~\cite{CMS:2019xai} on $\sigma(pp\to Z'\gamma\to q\bar{q}\gamma)$, where the analysis is performed without flavour tagging. As a result of including flavour tagging, our bounds are significantly stronger in comparison.

\begin{figure}[!tb]
\begin{center}
\subfloat[]{\label{fig:multijet}
\includegraphics[width=.45\textwidth]{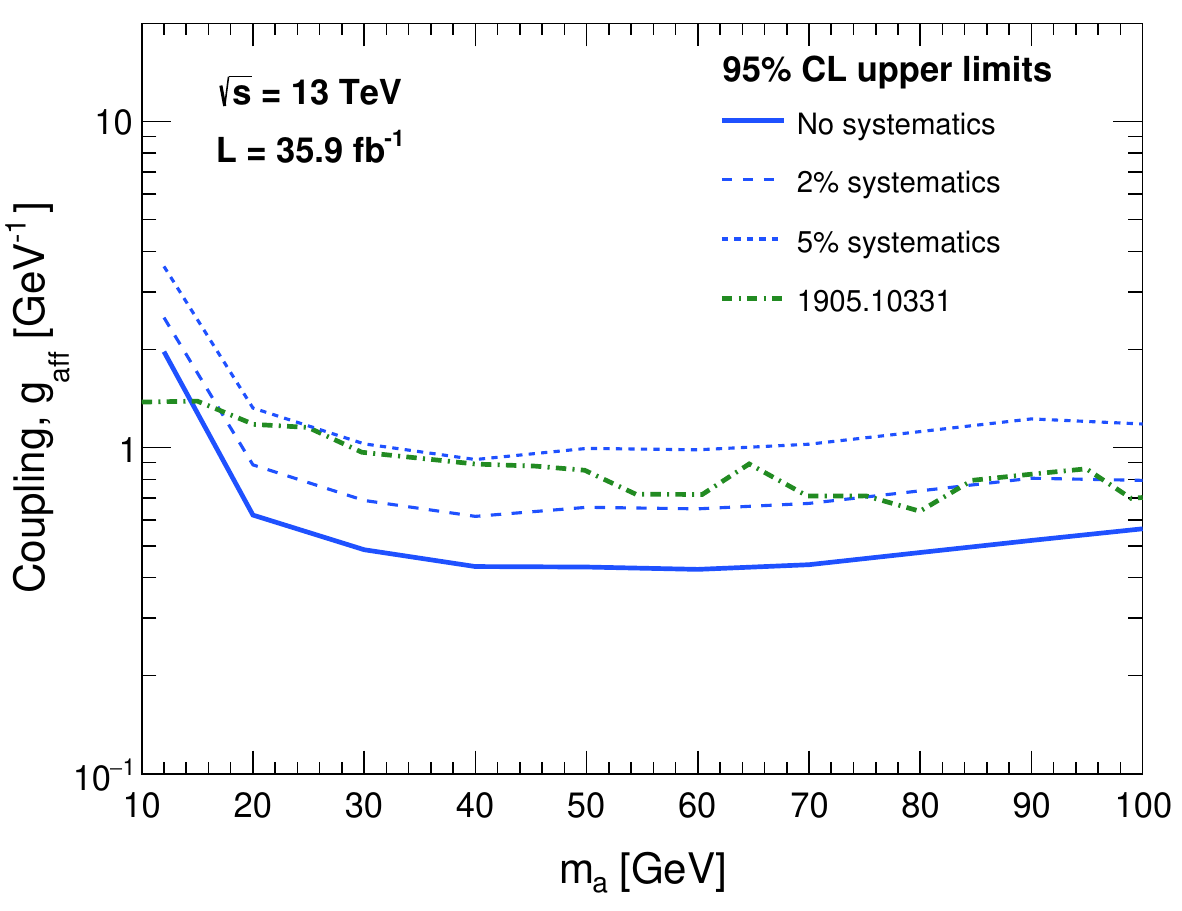}
}\qquad
\subfloat[]{\label{fig:alp}
\includegraphics[width=.45\textwidth]{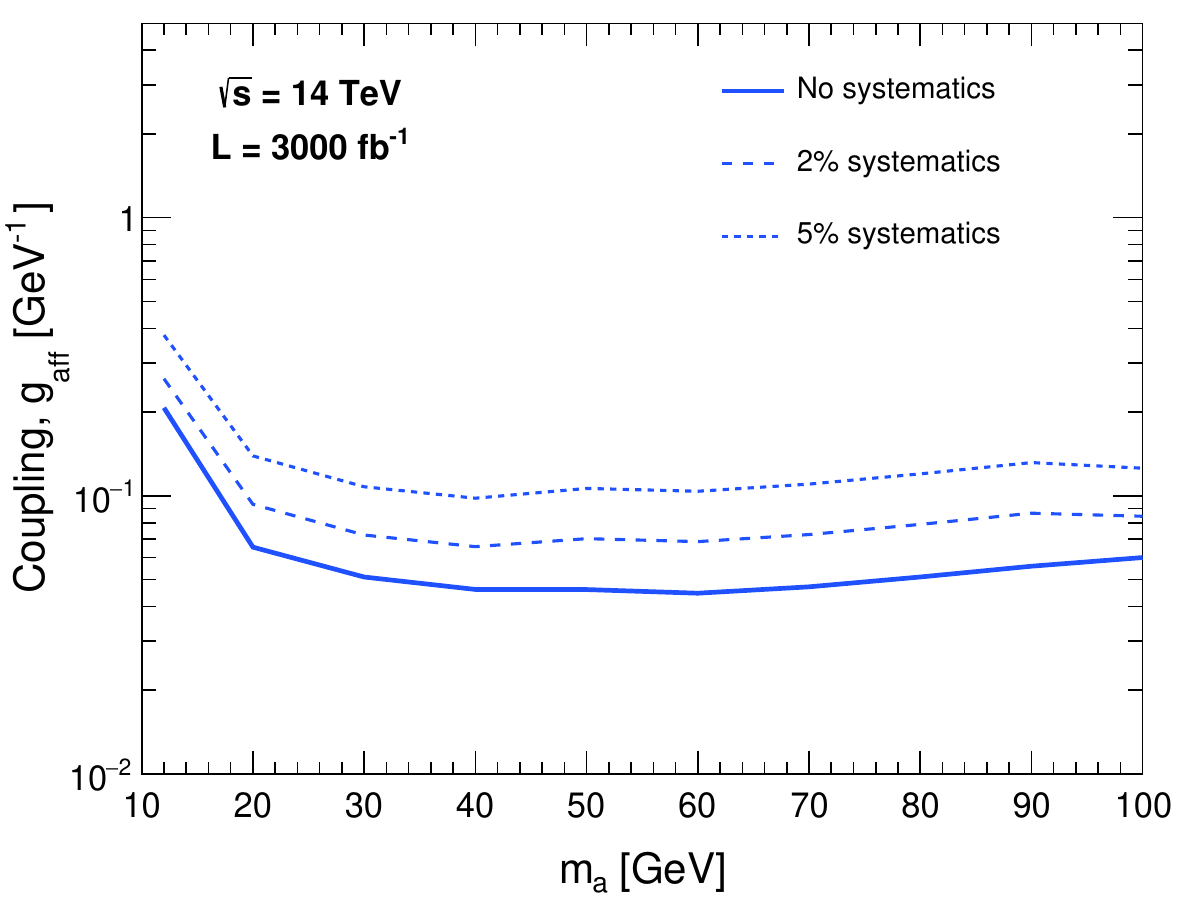}
}
\end{center}
\caption{Upper limit at $95\%$ C.L. on the ALP-fermion coupling $g_{\mathrm{aff}}$, as a function of the ALP mass. The limits are shown for (a) $\sqrt{s}=13$ TeV with $\mathcal{L}=35.9$ fb$^{-1}$ and (b) $\sqrt{s}=14$ TeV with $\mathcal{L}=3000$ fb$^{-1}$. The solid and (shorter) dashed blue lines correspond to neglecting systematic uncertainties or including them at the $2\%$ ($5\%$) level, respectively. The dashed green line correspond to the projected upper limits from Ref.~\cite{CMS:2019xai} }
\label{fig:UL_coupling}
\end{figure}

The derived upper limits on the signal production cross-section are translated into upper limits on the ALP-fermion coupling $g_{\mathrm{aff}}$, as illustrated as a function of $m_a$ in Fig.~\ref{fig:UL_coupling}. 
The left and right panel corresponds to upper limits at $\sqrt{s}=13$ TeV with $\mathcal{L}=35.9$ fb$^{-1}$ and $\sqrt{s}=14$ TeV with $\mathcal{L}=3000$ fb$^{-1}$, respectively. The couplings above the blue solid and (shorter) dashed line can be excluded at $95\%$ C.L. assuming zero or $2\%$ ($5\%$) systematic uncertainty, respectively. We observe that the stronger upper limit is obtained for an ALP mass $m_a\simeq 60$ GeV, with $g_{\mathrm{aff}}\le 0.045$ GeV$^{-1}$ at $\sqrt{s}=14$ TeV with $\mathcal{L}=3000$ fb$^{-1}$. For larger masses the limit slightly weakens, varying between $g_{\mathrm{aff}}\lesssim (0.047$ - $0.060$) GeV$^{-1}$ in the range $m_a\sim (70-100)$ GeV. Adding $2\%$ and $5\%$ systematic uncertainty, the coupling reach degrades to $g_{\mathrm{aff}}\lesssim (0.084$ - $0.26$) GeV$^{-1}$ and $(0.12$ - $0.38$) GeV$^{-1}$, respectively, in the ALP mass range between 12 GeV and 100 GeV.
We compare these limits with the results available at the current LHC run at $\sqrt{s}=13$ TeV with $\mathcal{L}=35.9$ fb$^{-1}$: the dashed green line is the projected $95\%$ C.L. upper limits on the coupling to SM quarks, in the searches for a $Z'$ resonance in association with a photon~\cite{CMS:2019xai}. We observe that our limits are stronger as compared to this reference, where the analyses were performed without any flavour decomposition of the jet.

\subsection{Analysis with low $p_{\rm T}$ threshold}
\label{sec:low_pho_pt}

\begin{figure}[!tb]
\begin{center}
\subfloat[]{\label{fig:multijet}
\includegraphics[width=.45\textwidth]{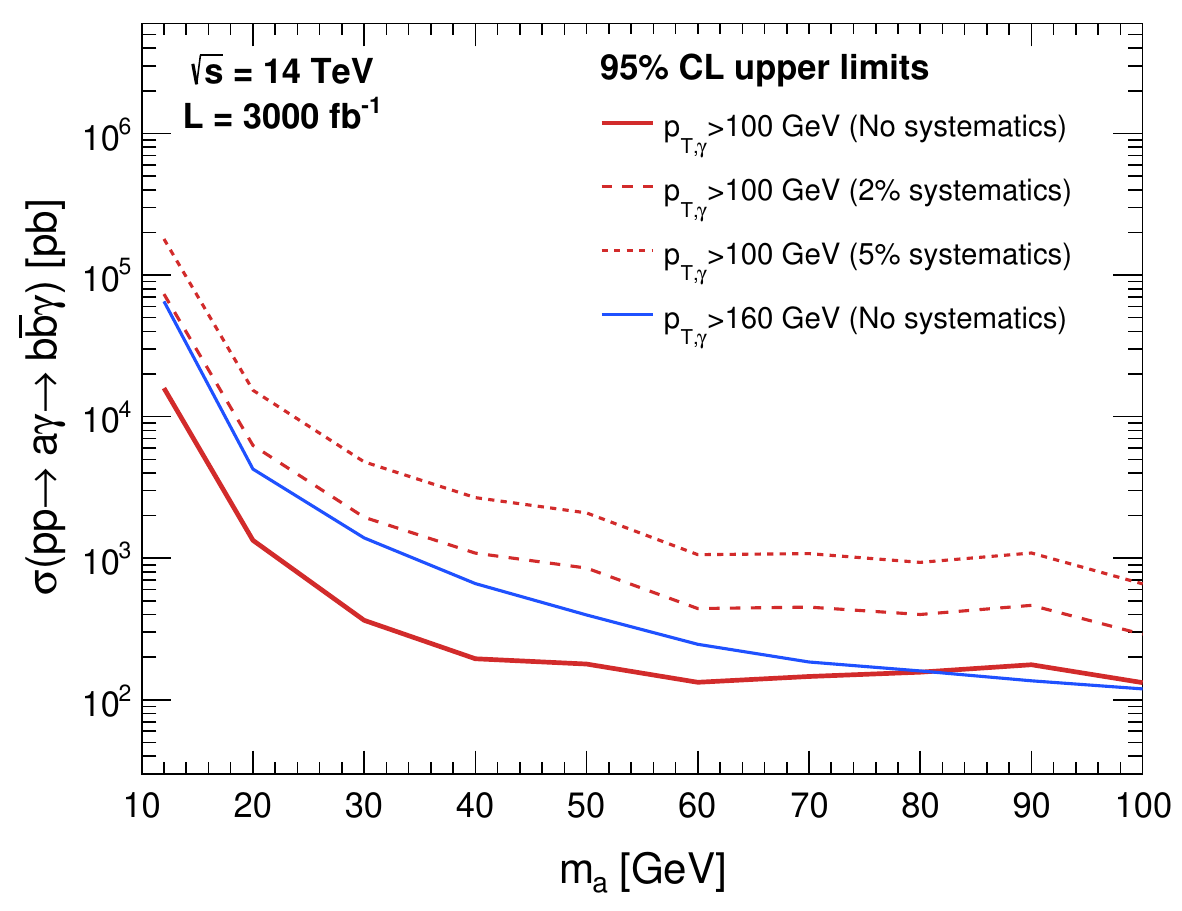}
}\qquad
\subfloat[]{\label{fig:alp}
\includegraphics[width=.45\textwidth]{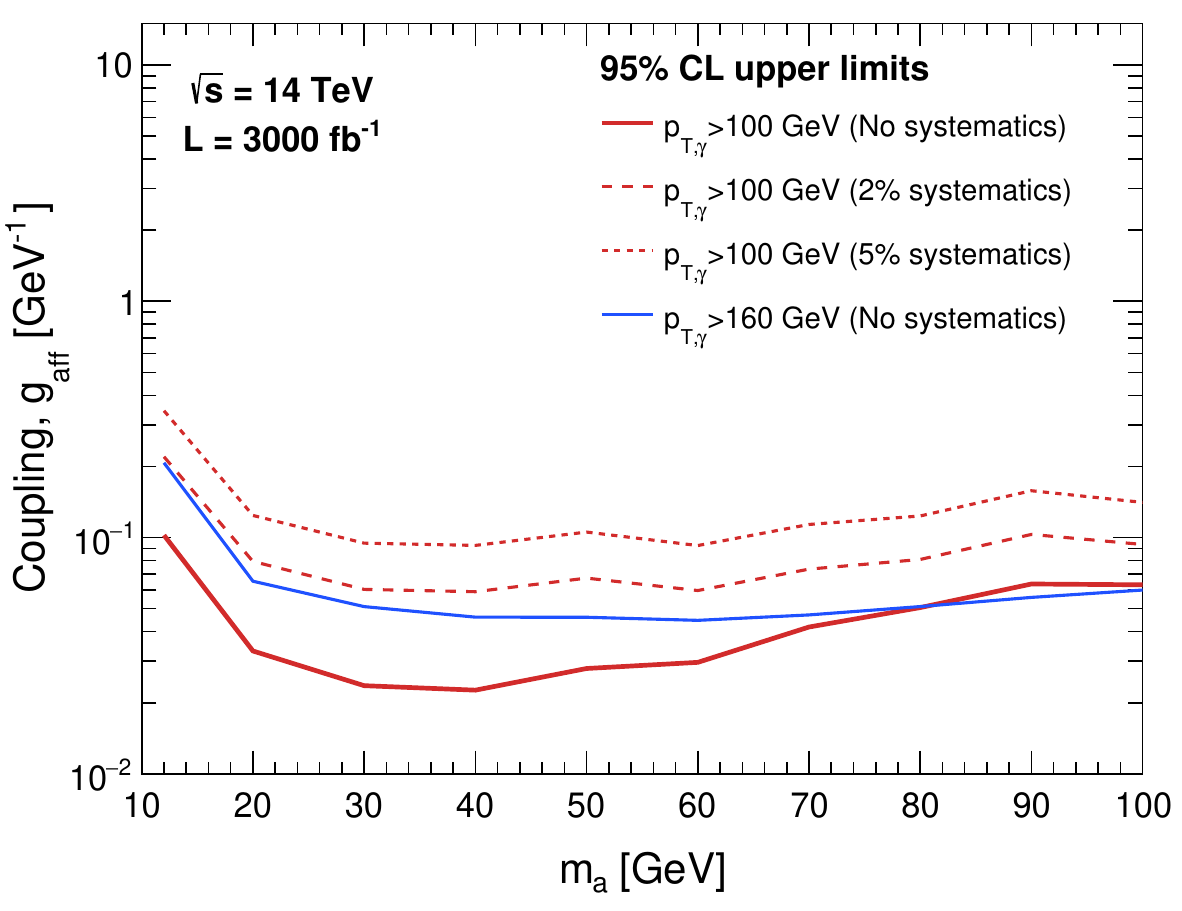}
}
\end{center}
\caption{ Upper limit at $95\%$ C.L. on (a) $\sigma(pp\to a\gamma\to b\bar{b}\gamma)$ and (b) $g_{\mathrm{aff}}$, with the requirement $p_{\rm T,\gamma}>100$ GeV, as a function of the ALP mass. The limits are shown for $\sqrt{s}=14$ TeV with $\mathcal{L}=3000$ fb$^{-1}$. Conventions are the same as in Fig.~\ref{fig:UL_sigma} and Fig.~\ref{fig:UL_coupling}. The limit using the earlier selection, $p_{\rm T,\gamma}>160$ GeV, is shown with blue solid line.}
\label{fig:UL_100}
\end{figure}

In the analysis presented earlier, we have adopted a $p_{\mathrm{T},\gamma}>160$ GeV threshold for the triggered photon, consistently with what is currently done by ATLAS~\cite{ATLAS:2018hbc}, discussed at the beginning of Section \ref{sec:event_sel}.
In what follows, we assess the impact on the expected sensitivity of lowering the photon threshold to $p_{\mathrm{T},\gamma}>100$ GeV,  by repeating the analysis described in Section~\ref{sec:collider} with looser requirements on the AK1.0 jet transverse momentum, i.e. $p_{\rm T}=100$ GeV. 
The $95\%$ C.L. upper limits on the ALP production cross-section, $\sigma(pp\to a\gamma\to b\bar{b}\gamma)$, and the ALP-fermion coupling, $g_{\mathrm{aff}}$ are shown in Fig.~\ref{fig:UL_100} and compared with the limits derived using the photon $p_{\mathrm{T},\gamma}>160$ GeV and $p_{\mathrm{T,AK1.0}}>200$ GeV trigger strategy. 

By loosing the requirement on the photon transverse momentum to $p_{\mathrm{T},\gamma}>100$ GeV, the cross-section, $\sigma(pp\to a\gamma\to b\bar{b}\gamma)$ limits improve by a factor four at $m_a=12$ GeV, varying between $1.6\cdot 10^4$ pb and $146$ pb for $m_a$ between $12$ and $70$ GeV (above $\sim 70$ GeV, the sensitivity remains similar to the previous selection, $p_{\mathrm{T},\gamma}>160$ GeV).  The ALP-fermion coupling can be probed down to $g_{\mathrm{aff}}\simeq 0.102$ GeV$^{-1}$ at $m_a=12$ GeV, while the maximal sensitivity, $g_{\mathrm{aff}}\simeq 0.023$ GeV$^{-1}$ is reached for $m_a\simeq 30-40$ GeV, although a slightly weaker bound is obtained
$\simeq 0.042$ GeV$^{-1}$ for $m_a\simeq 70$ GeV. Thus we find that the sensitivity to low-mass ALPs can be improved significantly with a stronger trigger requirement, in particular by lowering the $p_{\rm T}$ threshold for the accompanying photon.

\subsection{Interpretation of the results}
\label{sec:summ_result}

In order to assess the potential impact of these limits, it is interesting to visualise them along with existing constraints on the ALP parameter space. These existing constraints include limits from astrophysics, beam-dump and flavour experiments
for the region $m_a \lesssim 5$ GeV, as calculated in Ref.~\cite{Bharucha:2022lty} and ATLAS/CMS bounds for larger ALP masses, calculated here. More precisely, the $m_a$ - $g_{\rm aff}$ parameter space is constrained by:
astrophysical constraints from SN1987A~\cite{Kelly:2020dda,Ertas:2020xcc} and Horizontal Branch (HB) stars~\cite{Pospelov:2008jk,Avignone:1986vm}; electron beam dump experiments from SLAC E137~\cite{Bjorken:1988as} and heavy meson decays coming from LHCb, CHARM, NA62, NA48~\cite{LHCb:2016awg,LHCb:2015nkv,Dobrich:2018jyi,NA62:2021zjw,NA482:2009pfe,NA482:2016sfh}. 
For ALP masses $\gtrsim 5$ GeV , the bounds are derived from BSM searches at LHC, including dijet searches~\cite{CMS:2019xai,CMS:2017dcz}, di-photon searches~\cite{Mariotti:2017vtv,ATLAS:2022abz}, top-quark pair production with a lepton pair ($t\bar{t}\ell^+\ell^-$)~\cite{CMS:2024ulc}, 
and the $h\to$ BSM search~\cite{ATLAS:2022vkf}. In addition, SM measurements of $t\bar{t}$ production in association with a bottom pair ($t\bar{t}b\bar{b}$)~\cite{CMS:2020utv}, four-top-quark production ($t\bar{t}t\bar{t}$)~\cite{AlvarezGonzalez:2024tot,CMS:2023zdh}, and $t\bar{t}$ production~\cite{ATLAS:2019hxz} are used to constrain part of the model parameter space.

\begin{figure}[!tb]
\centering
\includegraphics[width=1\textwidth]{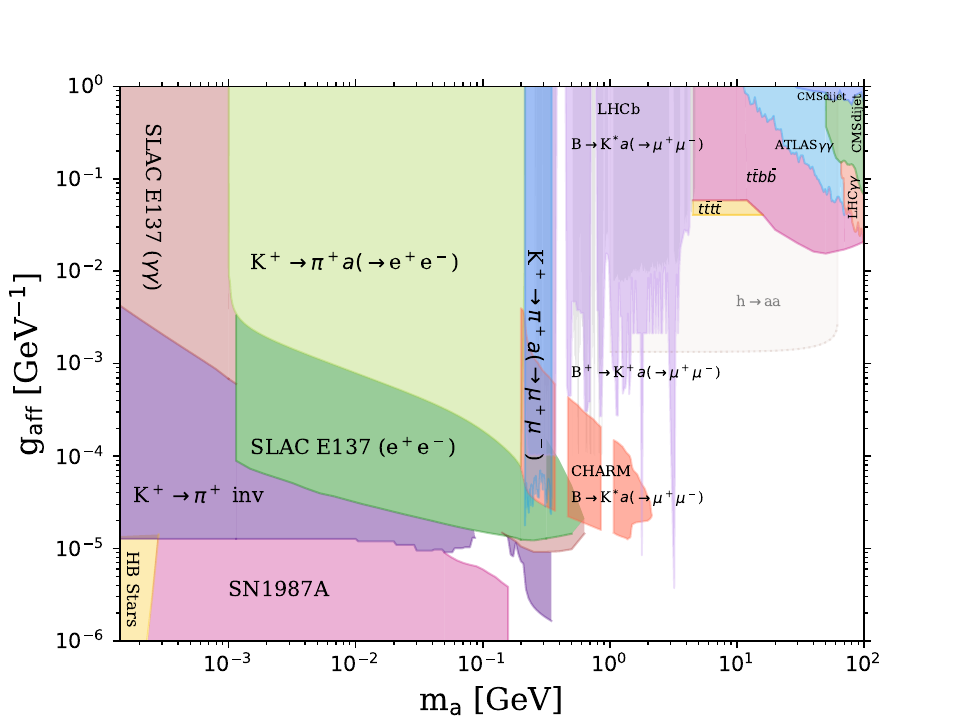} 
\caption{Constraints from astrophysical and collider experiments on the ALP-fermion coupling, $g_{\rm aff}$ as a function of the ALP mass, $m_a$. The astrophysical constraints include SN1987A~\cite{Kelly:2020dda,Ertas:2020xcc}, Horizontal Branch (HB) stars~\cite{Pospelov:2008jk,Avignone:1986vm}. The collider constrains come from electron beam dump experiments from SLAC E137~\cite{Bjorken:1988as}, heavy meson decays coming from LHCb, CHARM, NA62, NA48~\cite{LHCb:2016awg,LHCb:2015nkv,Dobrich:2018jyi,NA62:2021zjw,NA482:2009pfe,NA482:2016sfh}, di-photon searches~\cite{Mariotti:2017vtv,ATLAS:2022abz}, di-jet searches~\cite{CMS:2019xai,CMS:2017dcz}, $h\to$ BSM~\cite{ATLAS:2022vkf}, $t\bar{t}b\bar{b}$~\cite{CMS:2020utv} and $t\bar{t}t\bar{t}$~\cite{AlvarezGonzalez:2024tot,CMS:2023zdh} production.
}
\label{fig:constraints}
\end{figure}

In Fig.~\ref{fig:constraints}, we summarise the astrophysics and collider constraints that are adapted~\footnote{We are very grateful to Sophie Mutzel for providing the data required to reproduce these constraints.} from Ref.~\cite{Bharucha:2022lty}\, along with our derived collider bounds for larger ALP masses in the $g_{\rm aff}-m_a$ plane. The astrophysical bounds from SN1987A and HB stars exclude regions below $m_a\lesssim$ 100 MeV and $\sim$100 keV, respectively. The flavor  constrains for $m_a\lesssim 5$ GeV, exclude regions above $g_{\rm aff}\sim 10^{-5}$ GeV$^{-1}$. Above the ALP mass of $\sim 5$ GeV, we include the following collider constraints:
\begin{itemize}
\item The {\bf $\mathbf{h\to}$ BSM bound} arises as the SM Higgs can decay to ALPs via fermion loops and thus constrain the ALP-fermion coupling. We first calculate the $h\to aa$ branching ratio, considering the top- and bottom-quark loop contributions. We use FeynRules~\cite{Alloul:2013bka} to generate the model file for FeynArts~\cite{Hahn:2000kx}, then FeynArts, FormCalc and LoopTools~\cite{Hahn:1998yk} packages to generate and calculate the Feynman diagrams. We take a similar approach as in Ref.~\cite{Blasi:2023hvb}: we evaluate the scale-dependent loop diagrams, taken for the renormalisation scale $\mu=m_h$. 
An upper limit on the branching ratio $h\to$ BSM, Br($h\to$ BSM)$<0.073$, was derived in Ref.~\cite{Blasi:2023hvb} from ATLAS measurements~\cite{ATLAS:2022vkf}. This is effectively an upper bound on the derived Br($h\to aa$), as here this is the dominant BSM decay channel of the Higgs. We show this bound in Fig.~\ref{fig:constraints} in the $g_{\rm aff}-m_a$ plane. Notice however that tree-level contributions to $h\to aa$ are also possible due to the dim-6 operator $(\partial_\mu a)(\partial_\mu a
)H^\dagger H$, where the corresponding Wilson coefficient is independent of $g_{\rm aff}$. Therefore the $h\to$ BSM constraint can be relaxed by a cancellation between these two independent contributions and we hence show it by a dotted-shaded line in Fig.~\ref{fig:constraints}.
\item To constrain $g_{\rm aff}$ from the {\bf$\mathbf{t\bar{t}b\bar{b}}$ and $\mathbf{t\bar{t}t\bar{t}}$ processes}, we calculate the production cross-section in our ALP model using \texttt{MadGraph} and demand that they agree with the experimentally measured cross-section. In the case of the $t\bar{t}b\bar{b}$ process, the full phase space cross-section as defined in Ref.~\cite{CMS:2020utv} is evaluated with the following requirements on the jets: $p_{\rm T}>$ 20 GeV and $|\eta|<$ 2.4. An upper limit on $g_{\rm aff}$ is set by comparing the obtained cross-section to the measured value of 4.54 $\pm$ 0.35 (stat.) $\pm$ 0.56 (syst.) $\pm$ 0.66 (MG5) pb~\cite{CMS:2020utv}, as illustrated in Fig.~\ref{fig:constraints} by the magenta region. 
As the $t\bar{t}\ell^+\ell^-$ are similar to those from $t\bar{t}b\bar{b}$ (see Ref.~\cite{Blasi:2023hvb}), we only calculate the former. 
In the case of $t\bar{t}t\bar{t}$ production, the measured cross-section is 17 $\pm$ 4 (stat.) $\pm$ 3 (syst.) fb~\cite{AlvarezGonzalez:2024tot,CMS:2023zdh}. The corresponding upper bounds on $g_{\rm aff}$ from the $t\bar{t}t\bar{t}$ is shown in Fig.~\ref{fig:constraints} by the yellow region. 
\item  The {\bf dijet resonance searches} from CMS~\cite{CMS:2019xai,CMS:2017dcz} are shown by the blue and green regions in Fig.~\ref{fig:constraints}, similar to the left panel of Fig.~\ref{fig:UL_coupling}. In both ATLAS and CMS analyses~\cite{ATLAS:2022abz,Mariotti:2017vtv}, the {\bf di-photon constraints} are derived assuming a gluon fusion production mechanism, whereas in the ALP model considered here, the ALP production is initiated from quark annihilation. Thus, we reinterpret these limits by scaling the production cross-section times branching ratio, with the help of Ref.~\cite{Blasi:2023hvb}, and derive the bound in the $g_{\rm aff}-m_a$ plane, shown by the light-blue and red regions. 
\end{itemize}
Note that $t\bar{t}$ production could also exclude regions in the parameter space considered. However, we do not derive the bound here as the ALP contribution does not arise at tree level,  necessitating a calculation of the interference between the SM and BSM loop contribution~\cite{Maltoni:2024wyh} to reinterpret the bound in our model, which is beyond the scope of the present work.

\begin{figure}[!tb]
\centering
\includegraphics[width=1\textwidth]{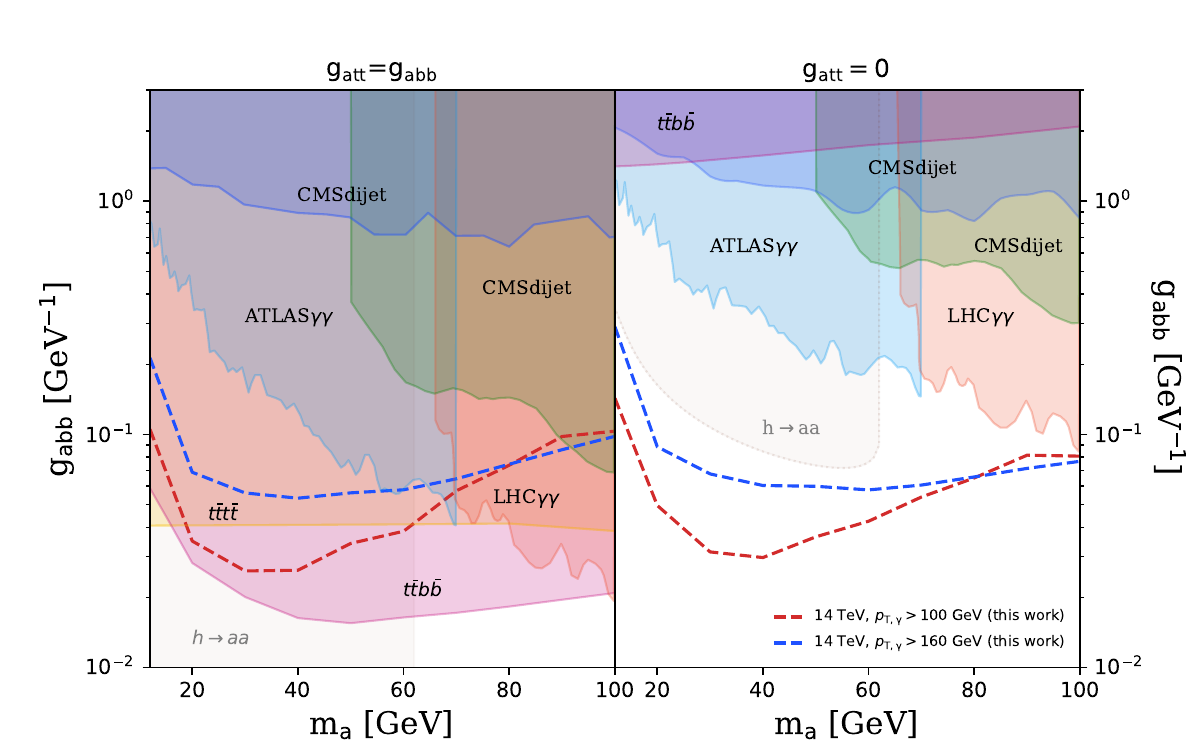}
\caption{The collider constraints on the ALP-fermion coupling, $g_{\rm aff}$ as a function of the ALP mass, $m_a$, in the mass range between $12-100$ GeV, is compared with our derived bounds. The collider constraints are the same as described in Fig.~\ref{fig:constraints}, while the derived bounds are shown with blue and red lines (as discussed in Fig.~\ref{fig:UL_100}). On the left panel, the bounds take into account the ALP couplings to top and bottom quark, while the ALP only couples to bottom quark for the constraints shown on the right panel. 
}
\label{fig:constraints2}
\end{figure}

While the ALP-top quark coupling has little relevance for our signal, it is crucial for several of the collider bounds mentioned above.
In Fig.~\ref{fig:constraints2}, we therefore focus on the region $10\lesssim m_a \lesssim 100$ GeV for the cases $g_{\rm att}=g_{\rm abb}$ (left panel) and $g_{\rm att}=0$ (right panel), in order to compare the existing constraints with our derived projected limits, shown by the blue and red dashed lines for $p_{\rm T,\gamma}>160$ GeV and $p_{\rm T,\gamma}>100$ GeV, respectively. The case of an ALP coupling to both the up-type and down-type quarks ($g_{\rm att}=g_{\rm abb}$), corresponding to Fig.~\ref{fig:constraints}, is illustrated in the left panel of Fig.~\ref{fig:constraints2}. 
We note that the constraint from $t\bar{t}b\bar{b}$ excludes a larger region than that from  $\gamma+b\bar{b}$ derived in this work. On the other hand, the $t\bar{t}t\bar{t}$ bound is flat over the considered mass range 
and weaker than the  $\gamma+b\bar{b}$ sensitivity for $p_{\rm T,\gamma}>100$ GeV, for ALP masses in the range 20-60 GeV. Below $m_a\sim 65$ GeV, the di-photon  constraint is weaker than the $\gamma+b\bar{b}$  bounds.

The right panel in Fig.~\ref{fig:constraints2} summarises the constraints on a class of models where the ALPs interact dominantly with down-type quarks ($g_{\rm att}=0$). 
As expected, all the previously discussed collider constraints become comparatively weaker than the $\gamma+b\bar{b}$ projected sensitivity for HL-LHC. 
In particular, for the $t\bar{t}b\bar{b}$ process, once we set the ALP-top coupling to zero, the only contributions are from quark-initiated diagrams, which involve top-quark emission from SM gauge or Higgs bosons. In the case of $h\to aa$, the bound only includes the bottom quark loop contribution. Similarly, the process $a\to \gamma\gamma$ is induced only by the bottom quark loop. It should also be noted that when the ALP-top coupling is zero, the ALP contribution to the $t\bar{t}t\bar{t}$ production vanishes. 
In contrast, our signal is only weakly affected when the ALP coupling to up-type quarks is switched off: the production cross-section is reduced roughly by a factor two, corresponding to the up-type quark content of the proton, and this relaxes the upper bound on $g_{\rm abb}$ by roughly a factor $\sqrt{2}$.

Other constraints on the ALP-top coupling have been investigated since Ref.~\cite{Bharucha:2022lty} appeared. In Ref.~\cite{Esser:2023fdo} the authors studied a scenario in which the only non-zero coupling of the ALP to SM fermions was that to top quarks, \textit{i.e} $g_{\rm att}\neq 0$ only. This has the important consequence that the ALP is long-lived with respect to the scale of the LHC detectors. As this is not the case in our model, the derived direct bounds do not apply here.
Indirect bounds from the LHC on the ALP-top coupling were also derived in Ref.~\cite{Esser:2023fdo}, and from low-energy, Higgs and top data on the ALP-fermion couplings in Ref.~\cite{Biekotter:2023mpd}. In the former, indirect effects of the ALP on top pair production and gauge-boson pair production are used to obtain limits on the ALP-top coupling. In the latter, bounds on the ALP couplings to up-type and down-type quarks as well as charged leptons were obtained by taking into account renormalization group flow of these couplings into the SM effective-field-theory (SMEFT) Wilson coefficients, on which the constraints were already known. It turns out that particularly strong bounds on couplings of ALPs to up-type were obtained, stemming from the two-loop contribution of the ALP-top coupling to the $W$ mass. While these bounds are clearly interesting, reproducing them for our model is beyond the scope of our work. We remark that the direct searches we consider provide important complementary information to these indirect bounds.

\section{Summary and outlook}
\label{sec:conclusion}

We have studied the LHC prospects for an ALP, a light pseudoscalar which primarily decays to the $b\bar{b}$ final state, in the mass range 10 to 100 GeV. We considered a simple DFSZ-inspired scenario where the ALP only couples to SM fermions and this coupling is proportional to the fermion Yukawa.

To enhance the sensitivity to low ALP masses, we required a recoiling, high transverse-momentum photon. Since the ALP is boosted, we employed a larger radius jet with $R=1.0$, to effectively capture the majority of the hadronisation products resulting from the decay of the ALP. Jet substructure techniques are implemented on the AK1.0 jet to reduce the multijet background. In this context, we utilised a two prong discriminant, $\mathrm{N}_2$, to identify the two prong structure in the signal AK1.0 jet, which originates from the $b\bar{b}$ final state. To reduce the $p_{\mathrm{T}}$ dependence of this observable, we modified it to a DDT observable,  $\mathrm{N}_2^{\mathrm{DDT}}$. However, the efficacy of this observable weakens for very low ALP masses, where the hadronisation products from the final state $b$-quarks merge and feature 1-prong substructure. 

A crucial aspect in our analysis involves the $b$-tagging of the AK1.0 jet. We performed it by constructing at least two variable radius subjets inside the AK1.0 jet and matched them to a hadron in the event, resulting in 6 possible di-flavour combinations for the multijet background: $bb$, $bc$, $bl$, $cc$, $cl$, and $ll$. Standard tagging efficiencies for single-jet flavour can not be applied here. We included the corresponding tagging efficiencies for the signal and multijet rejection to those di-flavour combinations for the background provided by the ATLAS collaboration~\cite{ATL-PHYS-PUB-2023-021}.

The final signal efficiency and background yields are calculated by performing a Crystal Ball fit to the signal ${\mathrm{m}}_{\mathrm{SD}}$  distribution, and selecting the $1\sigma$ window around the fitted mean value. The upper limits are set at $95\%$ confidence level on the $b\bar{b}\gamma$ production cross-section, $\sigma(pp\to a\gamma\to b\bar{b}\gamma)$. The HL-LHC limits vary between 185 pb and $4\cdot 10^{3}$ pb, without systematics, 440 (1014) pb and $8.7\cdot 10^{3}$ ($1.9\cdot 10^{4}$) pb, with $2\%$ ($5\%$) systematics, for ALP masses between 20 GeV and 70 GeV. Translating the cross-section limits to the ALP-fermion coupling, the upper bound on $g_{\mathrm{aff}}$, in the ALP mass range between 20 GeV and 70 GeV, is $(0.065$ - $0.047$) GeV$^{-1}$ at $95\%$ CL. The $g_{\mathrm{aff}}$ upper limits worsen upon adding systematic uncertainty, and the maximum obtainable reach is 0.066 (0.098) GeV$^{-1}$, around $m_a\sim$ 40 GeV, upon adding $2\%$ ($5\%$) systematics. 

Comparing these limits to the existing LHC searches at the current run, we find that our analysis provides significantly better exclusion limits on the $b\bar{b}\gamma$ production cross-section. In addition, we complement our analysis with a reduced  $p_{\mathrm{T},\gamma}$ threshold, to assess the capability of our analysis in increasing the sensitivity to low mass scenarios. Indeed, our study demonstrates that the sensitivity to the cross-section improves by a factor of four at $m_a=12$ GeV. The coupling $g_{\mathrm{aff}}$ can be excluded above $0.023$ GeV$^{-1}$ at $95\%$ CL for $m_a\sim 30-40$ GeV. 

We compare the results of this analysis to complementary existing astrophysical, flavor and collider searches. While the astrophysical and flavor bounds dominate around $m_a\lesssim m_b$, collider searches from the ATLAS and CMS experiment become important above an ALP mass of $m_a\gtrsim 10$ GeV. 
Comparing existing ATLAS and CMS constraints to our derived limits, the current bounds from $t\bar{t}b\bar{b}$ searches is slightly stronger than our bound.
However, these collider bounds heavily rely on the coupling of the ALP to the top quark. We find that, in models with no ALP coupling to up-type quarks, the analysis we propose would be the most powerful direct search for the ALP.

The analysis strategy presented can be readily applied to any well-motivated low mass resonance decaying into $b\bar{b}$.  We also demonstrated the improvement in sensitivity obtained by lowering the trigger thresholds. Moreover, analyses using machine learning techniques could further improve the reach over the traditional observables with jet substructure techniques. Future lepton colliders may also enhance the low mass resonance search prospects, as we plan to explore in future work.

\acknowledgments
We would like to thank Sophie Mutzel, Yann Coadou, Andreas Goudelis, Amandip De for helpful discussions, and Carlos Carranza for computing support, during the course of this work. MF wishes to thank the laboratories CPPM and CPT for hospitality during the completion of this work.
AA received support from the French government under the France 2030 investment plan, as part of the Initiative d'Excellence d'Aix-Marseille Université - A*MIDEX. MF received support from the European Union Horizon 2020 research and innovation program under the Marie Sk\l odowska-Curie grant agreements No 860881-HIDDeN and No 101086085–ASYMMETRY. This work received support from the French government under the France 2030 investment plan, as part of the Excellence Initiative of Aix Marseille University - amidex (AMX-19-IET-008 - IPhU)

\bibliographystyle{JHEP}
\bibliography{biblio}
\end{document}